\documentstyle[a4,boxedeps,cite]{article}
\newcommand{\G}{\gamma}
\newcommand{\GG}{{\gamma\gamma}}
\newcommand{\EPEM}{e^+e^-}
\newcommand{\MUPM}{\mu^+\mu^-}
\newcommand{\A}{\alpha}
\newcommand{\ZA}{Z\alpha}
\newcommand{\BE}{\begin{equation}}
\newcommand{\EE}{\end{equation}}

\makeatletter
\def\lesssim{\mathrel{\mathpalette\vereq<}}
\def\vereq#1#2{\lower3pt\vbox{\baselineskip1.5pt \lineskip1.5pt
\ialign{$\m@th#1\hfill##\hfil$\crcr#2\crcr\sim\crcr}}}
\def\gtrsim{\mathrel{\mathpalette\vereq>}}
\makeatother

\SetRokickiEPSFSpecial
\HideDisplacementBoxes

\begin{document}

\title{Photon-Photon Physics in Very Peripheral Collisions of
Relativistic Heavy Ions}

\author{Gerhard Baur\footnote{EMail:G.Baur@fz-juelich.de},\\
Institut f\"ur Kernphysik, Forschungszentrum J\"ulich, \\
Postfach 1913, D--52425 J\"ulich\\
Kai Hencken\footnote{EMail:hencken@quasar.physik.unibas.ch} and Dirk
Trautmann\footnote{EMail:trautmann@ubaclu.unibas.ch},\\
Institut f\"ur Physik, Universit\"at Basel,\\
Klingelbergstr. 82, CH--4056 Basel}

\date{\today \\
Topical Review, to appear in Journal of Physics G
}

\maketitle

\begin{abstract}
In central collisions at relativistic heavy ion colliders like the
Relativistic Heavy Ion Collider RHIC/Brookhaven and the Large Hadron
Collider LHC (in its heavy ion mode) at CERN/Geneva, one aims at
detecting a new form of hadronic matter --- the Quark Gluon Plasma. It
is the purpose of this review to discuss a complementary aspect of
these collisions, the very peripheral ones. Due to coherence, there
are strong electromagnetic fields of short duration in such
collisions. They give rise to photon-photon and photon-nucleus
collisions with high flux up to an invariant mass region hitherto
unexplored experimentally. After a general survey photon-photon
luminosities in relativistic heavy ion collisions are
discussed. Special care is taken to include the effects of strong
interactions and nuclear size. Then photon-photon physics at various
$\GG$-invariant mass scales is discussed.  The region of several GeV,
relevant for RHIC is dominated by QCD phenomena (meson and vector
meson pair production). Invariant masses of up to about 100 GeV can be
reached at LHC, and the potential for new physics is
discussed. Photonuclear reactions and other important background
effects, especially diffractive processes are also discussed.  A
special chapter is devoted to lepton-pair production,
especially electron-positron pair production; due to the strong fields
new phenomena, especially multiple $\EPEM$ pair production, will occur
there.

\end{abstract}

\tableofcontents

%%%%%%%%%%%%%%%%%%%%%%%%%%%%%%%%%%%%%%%%%%%%%%%%%%%%%%%%%%%%%%%%%%%%%%
\section{Introduction}

The parton picture is very useful to study scattering processes at
very high energies. In this model the scattering is described as an
incoherent superposition of the scattering of the various
constituents. For example, nuclei consist of nucleons which in turn
consist of quarks and gluons, photons consist of lepton pairs,
electrons consist of photons, etc. It is the subject of this topical
review to discuss that relativistic nuclei have photons as an
important constituent, especially for low enough virtuality
$Q^2=-q^2>0$ of the photon. This is due to the coherent action of all
the charges in the nucleus.  The virtuality of the photon is related
to the size $R$ of the nucleus by
\BE
Q^2 \lesssim 1/R^2, 
\EE
the condition for coherence. The radius of a nucleus is given
approximately by $R=1.2$~fm~$A^{1/3}$, where $A$ is the nucleon
number. From the kinematics of the process one has
\BE
Q^2=\frac{\omega^2}{\G^2}+Q_\perp^2
\EE
Due to the coherence condition the maximum energy of the
quasireal photon is therefore given by
\BE
\omega_{max} \approx \frac{\G}{R},
\label{eq_wmax}
\EE
and the maximum value of the perpendicular component is given by
\BE
Q_\perp \lesssim \frac{1}{R}.
\EE
We define the
ratio $x=\omega/E$, where $E$ denotes
the energy of the nucleus $E= M_N \G A$ and $M_N$ is the nucleon mass.
It is
therefore smaller than
\BE
 x_{max}=\frac{1}{R M_N A} = \frac{\lambda_C(A)}{R},
\EE
where $\lambda_C(A)$ is the Compton wave length of the ion. Here and
also throughout the rest of the paper we use natural units, setting
$\hbar=c=1$.

The collisions of $e^+$ and $e^-$ has been the traditional way to
study $\GG$-collisions. Similarly photon-photon collisions can also be
observed in hadron-hadron collisions. Since the photon number scales
with $Z^2$ ($Z$ being the charge number of the nucleus) such effects
can be particularly large. Of course, the strong interaction of the
two nuclei has to be taken into consideration.
%
% x-Definition Figure
%
\begin{figure}[tbhp]
\begin{center}
\ForceHeight{4cm}
\BoxedEPSF{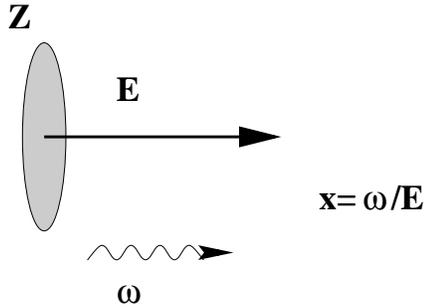}
\end{center}
\caption{\it
A fast moving nucleus with charge $Ze$ is surrounded by a strong
electromagnetic field. This can be viewed as a cloud of virtual
photons.  These photons can often be considered as real. They are
called equivalent or quasireal photons. The ratio of the photon energy
$\omega$ and the incident beam energy $E$ is denoted by $x=\omega/E$.
Its maximal value is restricted by the coherence condition to
$x<\lambda_C(A)/R\approx 0.175/A^{4/3}$, that is, $x\protect\lesssim
10^{-3}$ for Ca ions and $x\protect\lesssim 10^{-4}$ for Pb ions.  }

\label{fig_xvar}
\end{figure}
%
% AA gammagamma collisions
%
\begin{figure}[tbhp]
\begin{center}
\ForceHeight{5cm}
\BoxedEPSF{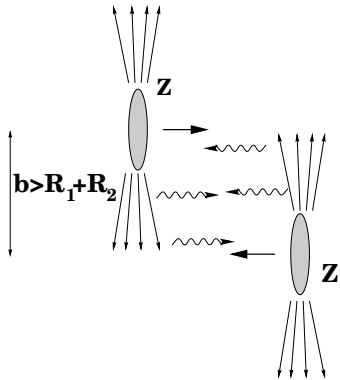}
\end{center}
\label{fig_collision}
\caption{\it
Two fast moving electrically charged objects are an abundant source of
(quasireal) photons. They can collide with each other and with the
other nucleus. For peripheral collisions with impact parameters $b>2R$,
this is useful for photon-photon as well as photon-nucleus
collisions.}
\end{figure}

The equivalent photon flux present in medium and high energy nuclear
collisions is very high, and has found many useful applications in
nuclear physics \cite{BertulaniB88}, nuclear astrophysics
\cite{BaurR94,BaurR96}, particle physics \cite{Primakoff51} (sometimes
called the ``Primakoff effect''), 
as well as, atomic physics \cite{Moshammer97}.
It is the main purpose of this review to discuss the physics of
photon-photon and photon-hadron (nucleus) collisions in high energy
heavy ion collisions. With the 
construction of the ``Relativistic Heavy Ion Collider'' (RHIC) and the
``Large Hadron Collider'' (LHC) scheduled for 1999 and for 2004/2008,
respectively, one will be able to investigate such collisions
experimentally. The main purpose of these heavy ion colliders 
is the formation and detection of the
quark-gluon-plasma, a new form of highly excited dense hadronic
matter. Such a state of matter will be created in central
collisions. The present interest is in the ``very peripheral (distant)
collisions'', where the nuclei do not interact strongly with each
other. From this point of view, grazing collisions and central
collisions are considered as a background. It is needless to say that
this ``background'' can also be interesting physics of its own.

The equivalent photon spectrum extends up to several GeV at RHIC
energies ($\G\approx 100$) and up to about 100 GeV at LHC energies
($\G\approx 3000$), see Eq.~(\ref{eq_wmax}).  Therefore the range of
invariant masses $M_{\GG}$ at RHIC will be up to about the mass of the
$\eta_c$, at LHC it will extend into an invariant mass range hitherto
unexplored.

We discuss the equivalent photon spectra of strongly interacting
particles, from which $\GG$-luminosities are obtained.  Due to the
coherence effect, the corresponding $\GG$-luminosity is very high.
In addition higher order and inelastic processes, which may occur in
heavy ion collisions are discussed.  Then 
the various possibilities for $\GG$-physics in the different invariant
mass regions will be explored.  A special case of $\GG$-physics is
lepton pair production, in particular the $e^+e^-$ pair
production. Since the equivalent photon approximation fails in some
regions of phase space, this case is discussed separately. Due to its
strong-field aspects, the production of multiple $\EPEM$-pairs is of
basic interest. Electron-positron pair creation is also of practical
interest due to the possibility of capturing an electron in a
K,L,\dots{} shell. This changes the charge state of the ion and leads
to a beam loss and thus to a decrease in the luminosity. The very
large cross section for the production of $\EPEM$ pairs (sometimes
called QED-electrons) is also an important background for detectors.

Relativistic heavy ion collisions have been suggested as a general
tool for two photon physics about a decade ago. Yet the study of a
special case, the production of $\EPEM$ pairs in nucleus-nucleus
collisions, goes back to the work of Landau and Lifschitz in 1934
\cite{LandauL34} (In those days, of course, one thought more about
high energy cosmic ray nuclei than relativistic heavy ion
colliders). In the meantime the importance of this process 
has become very clear, and many studies followed, e.g.,
\cite{Soff80,AlexanderGM87,BaurB88}. This subject will be dealt with
in detail in Sec.~\ref{chap_proc} and~\ref{chap_epem}, 
where also recent experimental results will be mentioned.

The general possibilities and characteristic features of two-photon
physics in relativistic heavy ion collisions have been discussed in
\cite{BaurB88}. The possibility to produce a Higgs boson via
$\GG$-fusion was suggested in \cite{GrabiakMG89,Papageorgiu89}. In
these papers the effect of strong absorption in heavy ion collisions
was not taken into account. This absorption is a feature, which is
quite different from the two-photon physics at $\EPEM$ colliders. The
problem of taking strong interactions into account was solved by using
impact parameter space methods in \cite{Baur90d,BaurF90,CahnJ90}. Thus
the calculation of $\GG$-luminosities in heavy ion collisions is put
on a firm basis and rather definite conclusions were reached by many
groups working in the field \cite{VidovicGB93}; for a recent review containing further
references see \cite{KraussGS97}. Subsequent studies --- to be
described in detail in this review --- revealed in a clear way that
the theoretical situation is basically understood.  This opens the way
for many interesting applications.
Up to now hadron-hadron collisions have not been used for
two-photon physics. An exception can be found in
\cite{Vannucci80}. There the production of $\mu^+\mu^-$ pairs at the
ISR was observed.  The special class of events was selected, where no
hadrons are seen associated with the muon pair in a large solid angle
vertex detector. In this way one makes sure that the hadrons do not
interact strongly with each other, i.e., one is dealing with
peripheral collisions (with impact parameters $b>2R$); the
photon-photon collisions manifest themselves as ``silent events''.

We feel that this is a very good basis for planning concrete
experiments, as it is done at RHIC
\cite{KleinS97a,KleinS97b,KleinS95a,KleinS95b} and LHC
\cite{HenckenKKS96,Felix97,BaurHTS98}.  This review aims at giving the
main physical ideas and provide the key formula and results.
Details can be found in the literature. A few new
results will also be presented. But the main emphasis is to discuss
the principle ideas and results in the field.  We hope that this
review will further stimulate future investigations. It is appropriate
to recall that RHIC will start operating in 1999, only a year from now.

%%%%%%%%%%%%%%%%%%%%%%%%%%%%%%%%%%%%%%%%%%%%%%%%%%%%%%%%%%%%%%%%%%%%%%
\section{General survey of peripheral collisions}
\label{sec_gensurv}

Let us first discuss the importance of electromagnetic interactions in
peripheral collisions for the case of elastic scattering.  The
strength of the Coulomb interaction is measured by the Coulomb (or
Sommerfeld) parameter $\eta$ which is given in terms of the nuclear
charges $Z_1$ and $Z_2$ by
\BE
\eta = \frac{Z_1 Z_2 e^2}{\hbar v}.
\EE
For (ultra)relativistic collisions we have $v\approx c$ and thus
$\eta\approx Z_1 Z_2 /137$. Therefore for $pp$-collisions we have always
$\eta\ll 1$ and the Born approximation, that is, one-photon
exchange, is applicable. For this case, i.e., for $\eta\approx 1/137$,
elastic scattering is reviewed, e.g., in \cite{BlockC85}.  (Experimental
result for $p\bar p$ scattering at $\sqrt{s}=546$ and~$1800$~GeV at
Fermilab are given in \cite{Abe94a,Abe94b,Abe94c}).  There is a
superposition of nuclear and Coulomb amplitudes and the elastic
differential scattering cross section is divided into three distinct
regions, separated by the value of the square of the momentum transfer
$t$. It is defined as usual as $t=(p_i-p_f)^2$, which is negative
(space-like momentum transfer) in our metric used. For
$|t|\ll|t|_{int}$ Coulomb scattering dominates, for $|t|\gg|t|_{int}$,
nuclear scattering dominates and for $|t| \approx |t|_{int}$ there is
Coulomb-nuclear interference. $|t|_{int}$ is given by (see Eq.~(3.13) of
\cite{BlockC85}, where the factor of $Z_1 Z_2$ is added for the
scattering of two nuclei with charge $Z_1$,$Z_2$ instead of two
protons): 
\BE
|t|_{int} \approx \frac{8\pi \A Z_1 Z_2}{\sigma_{tot}}.
\label{eq_tint}
\EE
The nuclear elastic scattering amplitude is usually parameterized
as
\BE
F_n = \frac{\left(\rho+i\right) \sigma_{tot}}{4\sqrt{\pi}} e^{B t/2},
\EE
where $\sigma_{tot}$ is the total cross section, $\rho=\frac{{\rm Re}
f_{c.m.}(0)}{{\rm Im} f_{c.m.}(0)}$ and $B$ is a slope parameter
related to the size of the hadron. 
In this normalization the differential cross section is given by
\BE
\frac{d\sigma}{dt} = \left| F_n \right|^2.
\EE

On the other hand for (very) heavy ions we have $\eta \gg 1$ and
semi-classical methods are appropriate to deal with elastic scattering
(see also \cite{Greider65}). For $\eta\gg1$ the Coulomb interaction is
very strong and the Born approximation is no longer valid. (In an
analogy with optics there is now Fraunhofer diffraction instead of
Fresnel diffraction). Instead one should use a Glauber approximation,
where the Coulomb interaction is taken into account to all orders. One
can say that many photons are typically exchanged in the elastic
collisions (in contrast to the Born approximation relevant for the
$pp$ case, where one photon exchange is sufficient).
One can now integrate over the impact parameter $b$ using the
saddle point approximation. Thus one recovers the classical picture of
scattering. The particles move essentially on a straight line with a
constant velocity and an impact parameter $b$.  For $Z_1=Z_2=Z$ at
grazing impact parameter $b=2R$ the momentum transfer is given by
(see, e.g., \cite{JacksonED})
\BE
|t|_{Coul} = \left(\Delta p\right)^2 =
\left(\frac{2Z^2e^2}{2Rv}\right)^2 \approx \frac{Z^4 e^4}{R^2 c^2}.
\EE
This momentum transfer determines the scattering angle $\theta=\Delta
p/p$, where $p$ is the momentum of the particle.

Let us compare this quantity with the corresponding one for
$pp$-scattering, Eq.~(\ref{eq_tint}).  With $\sigma_{tot} \approx 8
\pi R^2$ one has
\BE
|t|_{int} = \frac{\A Z^2}{R^2} ,
\label{eq_tint2}
\EE
to be compared to 
\BE
|t|_{Coul} = \frac{\A Z^2}{R^2} (\A Z^2).
\label{eq_tcoul}
\EE
It seems interesting to note that $|t|_{Coul}$ is $Z^2\A$ times the
corresponding quantity $|t|_{int}$. Evidently the corresponding
scattering angles are exceedingly small. For $\eta\approx 1$,i.e., for
$Z \approx 12$, the quantities $|t|_{int}$ and $|t|_{Coul}$
(Eq.~(\ref{eq_tint2}) and~(\ref{eq_tcoul})) are about equal, as should
be the case. For such values of $Z$ the change from Fresnel to
Fraunhofer scattering takes place. A more detailed discussion is given
in Ch.~5.3.4 of \cite{NoerenbergW80}, based on the work of
\cite{Frahn72}. 

\begin{figure}[tbhp]
\begin{center}
\ForceHeight{4cm}
\BoxedEPSF{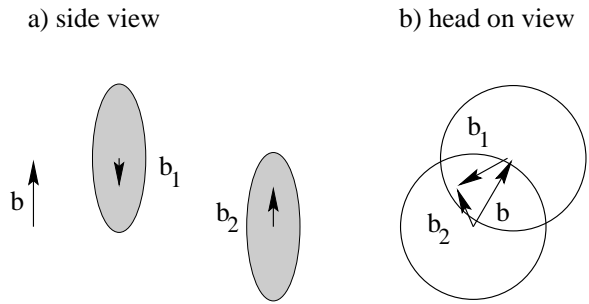}
\end{center}
\caption{\it
The probability of the nucleus to interact in central collisions can
be described within the $\rho$--$\rho$ folding approach to the Glauber 
theory. The parameters $b$, $b_1$ and $b_2$ used in the text are
explained.
}
\label{fig_central}
\end{figure}

The $pp$ cross section is rising with energy (see, e.g., p.193 of 
\cite{PDG96}). At LHC energies it will be of the order of 80~mb. What
does this mean for $p$-A and A-A cross section? A discussion is given
in \cite{KopeliovichNP88}. Experimental information can be obtained
from cosmic ray data for $p$-A reactions. Calculation for cross
section for A-A data often make use of a density-folding
approach. This approach can be justified starting from Glauber
theory\cite{Glauber67,EngelGLS92}.  One uses the thickness function
$T_A(b)$ of each nucleus, which is defined as the projection of the
density along the beam axis:
\BE
T_A(b)=\int_{-\infty}^{+\infty}
dz \ n_A\left(\sqrt{b^2+z^2}\right).
\label{eq_thickness}
\EE
Here the nuclear density $n_A(r)$ is normalized to $\int d^3r\
n_A(r)=A$.  From this one gets the ``overlap function'' given by
\BE
T_{AB}(b) = \int d^2b_1 \int d^2b_2 \ t(\vec b-\vec b_1 -\vec b_2)
T_A(b_1) T_B(b_2),
\label{eq_overlapf2}
\EE
where $\vec b$ is the impact parameter between the two ions (see also
Fig.~\ref{fig_central}).

$t(\vec s)$ describes the finite range of the nucleon-nucleon
interaction. It is proportional to the profile function $\Gamma(b)$ in
the Glauber theory, but it is normalized to $\int d^2s \ t(\vec
s)=1$. Therefore we have $\int d^2b T_{AB}(b) = A B$.  For high
energies the NN cross section rises beyond the geometrical size of the
protons. The finite range of the nucleon-nucleon interaction is
important then. A similar situation occurs also in nuclear physics at
lower energies, see, e.g., Figs.~2 and~3 of \cite{BertschBS89}, where
the influence of the finite range of the NN interaction is clearly
seen on the total cross-section for scattering of Li and Be isotopes.

If the nucleon-nucleon cross section is almost purely absorptive, the
scattering amplitude is almost imaginary and one can get $t(b)$ from
the elastic differential cross section
\BE
t(\vec s) = \frac{\int d^2p_\perp \exp(i \vec p_\perp \vec s)
\sqrt{\frac{d\sigma}{d^2p_\perp}}}{(2\pi)^2 
\left. \sqrt{\frac{d\sigma}{d^2p_\perp}}\right|_{p_\perp=0}}.
\EE
with $\frac{d\sigma}{d^2p_\perp}$ the elastic nucleon-nucleon
cross section.  Often
$t(\vec b)$ is approximated by an exponential function of the form
$t(\vec b) \sim \exp(-(b/b_0)^2)$, where $b_0$ is the range of the
interaction.

If one can neglect the finite range of the interaction compared to the
geometrical size of the nuclei, the ``thickness function'' can be
simplified:
\begin{eqnarray}
T_{AB}(b) &=& \int d^2b_1 \int d^2b_2 \delta^{(2)} (b-b_1 -b_2)
T_A(b_1) T_B(b_2) \\
&=& \int d^2b_1 T_A(b_1) T_B(b-b_1) .
\label{eq_overlapf1}
\end{eqnarray}

The inelastic scattering cross section of the nucleus-nucleus
collision is given by
\BE
\sigma_{AB,inel} = \int d^2b \left[1-\exp\left(- \sigma_{NN}
T_{AB}(b)\right)\right] ,
\EE
where $\sigma_{NN}$ is the total nucleon-nucleon cross
section. This integration over $b$ allows the integrand to be
interpreted as a probability for the two ions to interact. This
probability is about 1 for small impact parameters, where the
nuclei overlap, it drops to zero for large impact parameters. The
width of the area, where it falls off, is given by the surface
diffuseness of the nuclear densities and the range of the
nucleon-nucleon interaction as given by $t$.  Similarly the elastic
cross section is given by
\BE
\sigma_{AB,el} = \int d^2b \left[1-\exp\left(- \sigma_{NN}
T_{AB}(b)/2\right) \right]^2.
\EE

Using results from Regge theory (see, e.g., \cite{Abbiendi97,Engel97})
one can estimate the range $b_0$ of the interaction by
\BE
b_0 = 2 \sqrt{B_{0,pP}+\ln\left(s/s_0\right) \alpha_P'} .
\EE
Using for $B_{0,pP}\approx 2.4$GeV${}^{-2}$ and for
$\alpha_P'\approx0.25$GeV${}^{-2}$, as found in
\cite{Abbiendi97,Engel97}, one gets for LHC energies 
($s=(2\times7\mbox{TeV})^2$,$s_0=1\mbox{GeV}^2$) $b_0\approx 1$ fm,
which is small compared to the nuclear radius.

As we will see in the next section, the photon-photon luminosity
depends also on the probability for the two nuclei to interact with
each other. Only those processes, where the ions do not interact, are
useful for photon-photon physics. From our discussion here, we can
conclude that the effect of the increasing range of the NN
interaction will only be of some importance at the very high end of
the invariant $\GG$-masses for a quantitative
determination of the photon-photon luminosity.
%
% central collisions Figure
%

%%%%%%%%%%%%%%%%%%%%%%%%%%%%%%%%%%%%%%%%%%%%%%%%%%%%%%%%%%%%%%%%%%%%%%
\section{From impact-parameter dependent equivalent photon spectra to
{$\GG$-luminosities}}
\label{sec_lum}

\begin{figure}[tbhp]
\begin{center}
\ForceHeight{5cm}
\BoxedEPSF{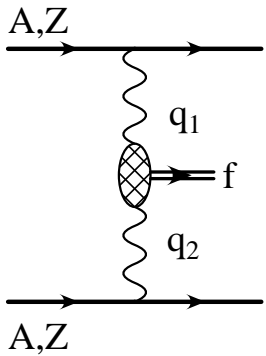}
\end{center}
\caption{\it
The general Feynman diagram of photon-photon processes in heavy ion
collisions: Two (virtual) photons fuse in a charged particle collision
into a final system $f$.
} 
\label{fig_ggcollision}
\end{figure}

Photon-photon collisions have been studied extensively at $\EPEM$
colliders. The theoretical framework is reviewed, e.g., in
\cite{BudnevGM75}.  The basic graph for the two-photon process in
ion-ion collisions is shown in Fig.~\ref{fig_ggcollision}. Two virtual
(space-like) photons collide to form a final state $f$. In the
equivalent photon approximation it is assumed that the square of the
4-momentum of the virtual photons is small, i.e., $q_1^2\approx
q_2^2\approx 0$ and the photons can be treated as quasireal. In this
case the $\GG$-production is factorized into an elementary cross
section for the process $\G+\G\rightarrow f$ (with real photons, i.e.,
$q^2=0$) and a $\GG$-luminosity function. In contrast to the pointlike
elementary electrons (positrons), nuclei are extended, strongly
interacting objects with internal structure. This gives rise to
modifications in the theoretical treatment of two photon processes.

\subsection{Elastic vertices}

The emission of a photon depends on the (elastic) form factor. Often a
Gaussian form factor or one of a homogeneous charged sphere is used.
The typical behavior of a form factor is
\BE
f(q^2) \approx 
\left\{
 \begin{array}{lcl}
 Z &\qquad& \mbox{for $|q^2| < \frac{1}{R^2}$}\\
 0 &\qquad& \mbox{for $|q^2| \gg \frac{1}{R^2}$}
 \end{array}
\right. .
\EE
For low $|q^2|$ all the protons inside the nucleus act coherently,
whereas for $|q^2| \gg 1/R^2$ the form factor is very small, close to
0. For a medium size nucleus with, say, $R=5$ fm, the limiting
$Q^2=-q^2=1/R^2$ is given by $Q^2=(40$MeV$)^2=1.6\times
10^{-3}$~GeV${}^2$. Apart from $\EPEM$ (and to a certain extent also
$\mu^+\mu^-$) pair production --- which will be discussed separately
below --- this scale is much smaller than typical scales in the
two-photon processes. Therefore the virtual photons
in relativistic heavy ion collisions can be treated as quasireal. This
is a limitation as compared to $\EPEM$ collisions, where the
two-photon processes can also be studied as a function of the
corresponding masses $q_1^2$ and $q_2^2$ of the exchanged photon
(``tagged mode'').
%
% Figure General gg-process
%
%
% Figure elastic emission
%
\begin{figure}[tbhp]
\begin{center}
\ForceHeight{4cm}
\BoxedEPSF{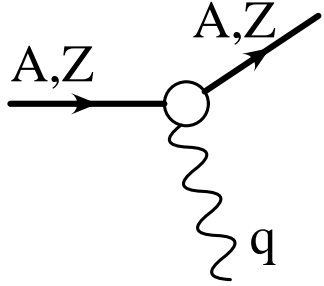}
\end{center}
\caption{\it
``Elastic photon emission'': In order for the photon to interact
coherently with the whole nucleus, its virtuality $Q^2$ is restricted
to $Q^2 \protect\lesssim 1/R^2$.  In the calculation this is
incorporated by the elastic form factor $f(q^2)$. This is important
for a treatment, where plane waves are used. In the semiclassical (or
Glauber) method, the detailed form of $f(q^2)$ is not relevant for
collisions with $b>R$ (see text).
} 
\label{fig_ggemission}
\end{figure}
%
% Figure elastic emission
%
\begin{figure}[tbhp]
\begin{center}
\ForceHeight{4cm}
\BoxedEPSF{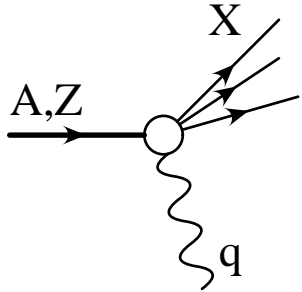}
\end{center}
\caption{\it 
``Inelastic photon emission'': As a nucleus is a rather weakly bound
system, photon emission can lead to its breakup or excitation,
especially for $Q^2 \protect\gtrsim 1/R^2$. Also for this incoherent
case an equivalent photon number, similar to the coherent case, can be
defined.
} 
\label{fig_ggemincoherent}
\end{figure}

As was discussed already in the previous section, relativistic
heavy ions interact strongly when the impact parameter is smaller than
the sum of the radii of the two nuclei. In such cases $\GG$-processes
are still present and are a background that has to be considered in
central collisions. In order to study ``clean'' photon-photon events
however, they have to be eliminated in the calculation of
photon-photon luminosities as the particle production due to the
strong interaction dominates. In the usual treatment of photon-photon
processes in $\EPEM$ collisions plane waves are used and there is
no direct information on the impact parameter. For heavy ion
collisions on the other hand it is very appropriate to introduce
impact parameter dependent equivalent photon numbers. They have been
widely discussed in the literature (see
e.g. \cite{BertulaniB88,JacksonED,WintherA79})

The equivalent photon spectrum corresponding to a point charge $Z e$,
moving with a velocity $v$ at impact parameter $b$ is given by
\BE
N(\omega,b) = \frac{Z^2\A}{\pi^2} \frac{1}{b^2}
\left(\frac{c}{v}\right)^2 x^2 \left[ K_1^2(x) + \frac{1}{\G^2}
K_0^2(x)\right],
\label{eq_nomegab}
\EE
where $K_n(x)$ are the modified Bessel Functions (MacDonald
Functions) \cite{AbramowitzS64} and $x=\frac{\omega b}{\G v}$. 
Then one obtains the probability for a certain electromagnetic process
to occur in terms of the same process generated by an equivalent pulse
of light as
\BE
P(b) = \int \frac{d\omega}{\omega} N(\omega,b) \sigma_\G(\omega).
\EE
Possible modifications of $N(\omega,b)$ due to an
extended spherically symmetric charge distribution are given in
\cite{BaurF91} (see also Eq.~(\ref{eq_nwb}) below).  It should be
noted that Eq.~(\ref{eq_nomegab}) also describes the equivalent photon
spectrum of an extended charge distribution, such as a nucleus, as long
as $b$ is larger than the extension of the object. This is due to the
fact that the electric field of a spherically symmetric system depends
only on the total charge, which is inside it. As one often wants to
avoid also final state interaction between the produced system and the
nuclei, one has to restrict oneself to $b_i>R_i$ and therefore
the form factor is not very important. For inelastic vertices a photon
number $N(\omega,b)$ can also be defined, as will be discussed in
Sec.~\ref{sec_incoherent} below.
%
% field of one ion Figure
%
\begin{figure}[tbhp]
\begin{center}
\ForceHeight{4cm}
\BoxedEPSF{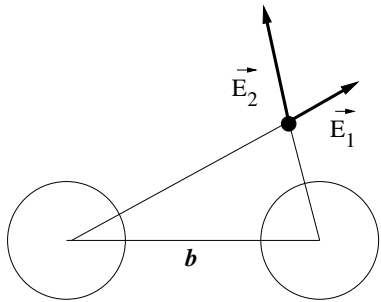}
\end{center}
\caption{\it
View of the collision perpendicular to the beam direction: The
electric field vector points along the direction of the individual
impact parameter.
}
\label{fig_4}
\end{figure}

As the term $x^2 \left[ K_1^2(x) + 1/\G^2 K_0^2(x)\right]$ in
Eq.~(\ref{eq_nomegab}) can be roughly approximated as 1 for $x<1$ and
0 for $x>1$, so that the equivalent photon number
$N(\omega,b)$ is almost a constant up to a maximum
$\omega_{max}=\G/b$ ($x=1$). By integrating
the photon spectrum (Eq.~(\ref{eq_nomegab}))
over $b$ from a minimum value of $R_{min}$ up to infinity (where
essentially only impact parameter up to  $b_{max}\approx \G/\omega$ 
contribute, compare with Eq.~(\ref{eq_wmax})),
one can define an equivalent photon number
$n(\omega)$. This integral can be carried out analytically and is
given by \cite{BertulaniB88,JacksonED}
\BE
n(\omega) = \int d^2b N(\omega,b) = \frac{2}{\pi} Z_1^2 \alpha
\left(\frac{c}{v}\right)^2 \left[ \xi K_0 K_1 - \frac{v^2\xi^2}{2 c^2}
\left(K_1^2 - K_0^2\right)\right] ,
\label{eq_nomegaex}
\EE
where the argument of the modified Bessel functions is 
$\xi=\frac{\omega R_{min}}{\G v}$.
The cross section for a certain electromagnetic process is then
\BE
\sigma = \int \frac{d\omega}{\omega} n(\omega) \sigma_{\G}(\omega).
\label{eq_sigmac}
\EE
Using the approximation above for the MacDonald functions, we get
an approximated form, which is quite reasonable and is useful for
estimates:
\BE
n(\omega) \approx  \frac{2 Z^2 \A}{\pi} \ln
\frac{\G}{\omega R_{min}}.
\label{eq_nomegaapprox}
\EE

The photon-photon production cross-section is obtained in a similar
factorized form,
by folding the corresponding equivalent photon spectra of the two
colliding heavy ions \cite{BaurF90,CahnJ90} (for
polarization effects see \cite{BaurF90}, they are neglected here)
\BE
\sigma_c = \int \frac{d\omega_1}{\omega_1} \int
\frac{d\omega_2}{\omega_2}
F(\omega_1,\omega_2) \sigma_{\GG}(W_{\GG}=\sqrt{4 \omega_1\omega_2}) ,
\label{eq_sigmaAA}
\EE
with
\begin{eqnarray}
F(\omega_1,\omega_2)&=& 2\pi \int_{R_1}^{\infty} b_1 db_1 
\int_{R_2}^{\infty} b_2 db_2 \int_0^{2\pi} d\phi \nonumber\\
&&\times N(\omega_1,b_1) N(\omega_2,b_2)
\Theta\left(b_1^2+b_2^2-2 b_1 b_2
\cos\phi-R_{cutoff}^2\right) ,
\label{eq_fw1w2}
\end{eqnarray}
($R_{cutoff} = R_1 + R_2$).  This can also be rewritten in terms of
the invariant mass $W_{\GG}=\sqrt{4\omega_1\omega_2}$ and the rapidity
$Y=1/2 \ln((P_0+P_z)/(P_0-P_z))=1/2 \ln(\omega_1/\omega_2)$ as:
\BE
\sigma_c = \int dW_{\GG} dY \frac{d^2L}{dW_{\GG} dY}
\sigma_{\GG}(W_{\GG}) ,
\label{eq_sigmaAAMY}
\EE
with 
\BE
\frac{d^2L_{\GG}}{dW_{\GG} dY} = \frac{2}{W_{\GG}}
F\left(\frac{W_{\GG}}{2} e^Y,\frac{W_{\GG}}{2} e^{-Y}\right) .
\label{eq_dldwdy}
\EE
Here energy and momentum in the beam direction are
denoted by $P_0$ and $P_z$. The transverse momentum is of the order of
$P_\perp \le 1/R$ and is neglected here. The transverse momentum
distribution is calculated in \cite{BaurB93}.

In \cite{BaurB93} and \cite{Baur92}
this intuitively plausible formula is derived ab
initio, starting from the assumption that the two ions move on a
straight line with impact parameter $b$. Eqs.~(\ref{eq_sigmaAA})
and~(\ref{eq_sigmaAAMY}) are the basic formulae for $\GG$-physics in
relativistic heavy-ion collisions. The advantage of heavy nuclei is
seen in the coherence factor $Z_1^2 Z_2^2$ contained in
Eqs.~(\ref{eq_sigmaAA})--(\ref{eq_dldwdy}).

Let us make a few remarks: In Eq.~(\ref{eq_fw1w2}) a sharp cut-off in
impact parameter space at $b_{min}=R_{cutoff}$ is introduced. There is
some ambiguity in the numerical value of $R_{cutoff}$: As was
discussed in Sec.~\ref{sec_gensurv} 
the total cross section for nucleon-nucleon collisions is
rising with energy. To a certain extent this will also affect the
values of $R_{cutoff}$. They are not just the (energy independent)
radii of the nuclei, but they rather describe the probability of
nuclear interactions between the heavy ions. A more realistic
calculation can be done by replacing this sharp cutoff with a smooth
one, using the overlap function of Eq.~(\ref{eq_overlapf1})
and~(\ref{eq_overlapf2}). The $\Theta$-function in
Eq.~(\ref{eq_fw1w2}) has to be replaced by
\BE
P(b) = 1- \exp\left(-\sigma_{NN} T_{AB}(b)\right) ,
\EE
with $\vec b = \vec b_1 - \vec b_2$.  Comparing this refined model
with the one with a sharp cutoff, significant deviations are only
present at the very upper end of the invariant mass range. Only the
smallest impact parameter contribute significantly to these events,
therefore a sensitivity to the cutoff is expected. All other (smaller)
invariant masses are not very sensitive to the details of this cutoff.

\begin{figure}[tbhp]
\begin{center}
\ForceHeight{7cm}
\BoxedEPSF{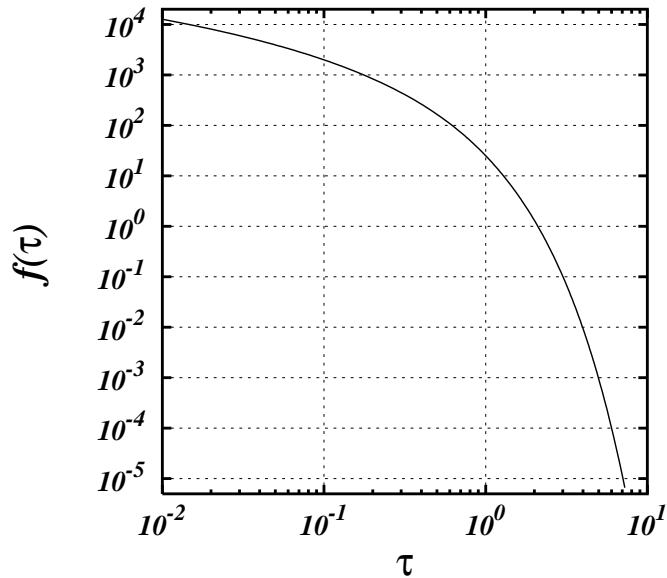}
\end{center}
\caption{\it
Plot of the universal function $f(\tau)$ (see
Eq.~(\protect\ref{eq_ftau})) as a function of $\tau=W_{\GG} R/\G$.
}
\label{fig_zuniversal}
\end{figure}

Integrating over all $Y$ the cross section is
\BE
\sigma_c = \int dW_{\GG} \frac{dL}{dW_{\GG}} 
\sigma_{\GG}(W_{\GG}).
\EE
For symmetric collisions ($R_1=R_2=R$) we can write the luminosity in
terms of a universal function $f(\tau)$ as
\BE
dL/dW_{\GG}=\frac{Z^4 \A^2}{\pi^4} \frac{1}{W_{\GG}} f(\tau),
\EE 
with $\tau=W_{\GG} R/\G$ \cite{CahnJ90}.
For large values of $\G$, $f(\tau)$ is given by 
\begin{eqnarray}
f(\tau) &=& \pi \tau^2 \int_{1}^{\infty} u_1 du_1 
\int_{1}^{\infty} u_2 du_2
\int_0^{2\pi} d\phi \ \Theta(u_1^2+u_2^2-2 u_1 u_2 \cos\phi - 4)
\nonumber\\ &&
\int_{-\infty}^{+\infty} dY K_1^2(\frac{\tau}{2} u_1 e^Y)
K_1^2(\frac{\tau}{2} u_2 e^{-Y}) .
\label{eq_ftau}
\end{eqnarray}
The function $f(\tau)$ is shown in 
Fig.~\ref{fig_zuniversal}, a useful parameterization of it was studied in 
\cite{CahnJ90}. 
\begin{table}[tbhp]
\begin{center}
\begin{tabular}{|c|c|c|c|c|}
\hline
Ion & $Z$,$A$ & $\sqrt{s}$ & $\G$ & $L_{AA}$ (cm${}^{-2}$ s${}^{-1}$)
\\
\hline
\multicolumn{5}{|c|}{LHC}\\
\hline
Pb-Pb & 82,208 & 1148 TeV & 2950 & $1 \times 10^{26}$\\
Ca-Ca & 20,40 & 280 TeV & 3750 & $4 \times 10^{30}$\\
$p$-$p$ & 1,1 & 14 TeV & 7450 & $10^{29}-10^{31}$\\
\hline
\multicolumn{5}{|c|}{RHIC}\\
\hline
Au-Au & 79,197 & 20 TeV & 106 & $2 \times 10^{26}$\\
I-I   & 53,127 & 13 TeV & 111 & $3 \times 10^{27}$\\
Cu-Cu & 29,63  & 7.2 TeV& 122 & $1 \times 10^{28}$\\
Si-Si & 14,28  & 3.5 TeV& 133 & $4 \times 10^{28}$\\
O-O   & 8,16   & 2.0 TeV& 133 & $1 \times 10^{29}$\\
$p$-$p$& 1,1   & 250 GeV& 266 & $1 \times 10^{31}$\\
\hline
\end{tabular}
\caption{\it
Parameters for different ion species at RHIC and LHC. In the entries we
give the total invariant mass $\sqrt{s}$ of the system, the Lorentz
factor $\G$ and the beam luminosity. Parameters are taken from 
\protect\cite{Felix97,KleinS95b}. 
}
\label{tab_lum}
\end{center}
\end{table}

As a function of $Y$, the luminosity $d^2L/dW_{\GG}{dY}$ has a
Gaussian shape with the maximum at $Y=0$. The width is approximately
given by $\Delta Y = 2 \ln \left[(2\G)/(R W_{\GG})\right]$. Depending
on the experimental situation additional cuts in the allowed $Y$ range 
are needed.

%%%%%%%%%%%%%%%%%%%%%%%%%%%%%%%%%%%%%%%%%%%%%%%%%%%%%%%%%%%%%%%
\subsection{Inelastic Vertices and Higher Order Corrections}
\label{sec_incoherent}

Heavy ions are complex objects
unlike the pointlike, structureless electrons. The effects due to the
finite size (and the nuclear interactions between them) has been
considered by using an impact parameter approach. The additional
processes coming from elastic nuclear interactions (diffractive
processes, Pomeron interactions) will be considered separately in
Sec.~\ref{sec_ga} and~\ref{sec_diffraction}. Here we want to discuss
mainly additional electromagnetic processes that can occur in distant
collisions. 

Especially for very heavy ions --- like Pb --- $\ZA$ is no longer
small. Therefore processes with more than one photon-exchange are
important.  In addition to the $\GG$-process one can have
electromagnetic dissociation of the ions. Furthermore the ions can
also be excited due to the emission of the photon. This excitation can
lead to an excited nucleus, or to the breakup of the nucleus, when the
proton is knocked out of the nucleus due to the photon emission. At
even higher $Q^2$ the photon can be emitted also from the quarks
contained in the protons.

In Sec.~\ref{sec_ga} we will see that the electromagnetic dissociation
of the ions is an important process. This cross section is often so
large that it can also occur in addition to the $\GG$-process.
%
% FD for GDR coherent + photon-photon
%
\begin{figure}[tbhp]
\begin{center}
\ForceWidth{0.6\hsize}
\BoxedEPSF{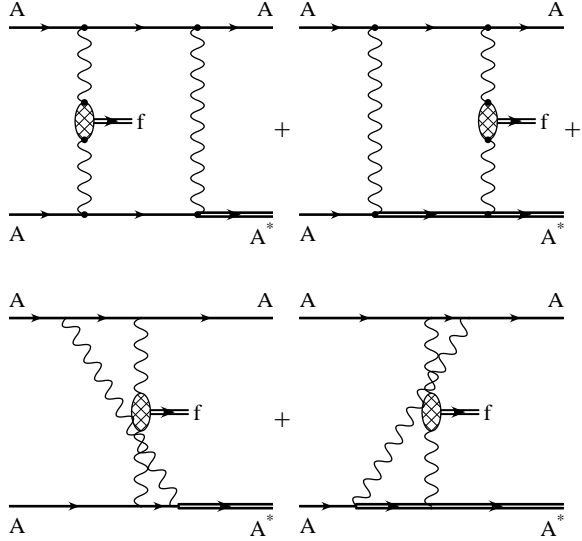}
\end{center}
\caption{\it
Due to the large charge of the ions, electromagnetic excitation in
addition to the photon-photon process may occur. 
}
\label{fig_gdrcoherent}
\end{figure}

In the impact parameter dependent approach the probability of several
processes to occur in one collision can be calculated by assuming that
the processes are independent of each other; their probabilities have
to multiplied:
\BE
P_{fA^*}(b) = P_{\GG \rightarrow f} (b) \ P_{\G A 
\rightarrow A^*} (b).
\EE
Therefore the processes given in Fig.~\ref{fig_gdrcoherent} are
effectively included in this approach.  This is a good approximation
as long as the nucleus is not disturbed substantially.  Integrating
from $b=2R$ up to infinity, one obtains the cross section for
$\GG \rightarrow f$ fusion accompanied by a specific $\G
A \rightarrow A^*$ interaction
\BE
\sigma_{fA^*} = 2 \pi \int_{2R}^{\infty} b db \ 
  P_{\GG \rightarrow f} (b) 
  \ P_{\G A \rightarrow A^*} (b).
\label{Eq:sigmafA*}
\EE

Since $\sum_{A^*} P_{\G A \rightarrow A^*} = 1$ (where we sum over
all possible states $A^*$ of the nucleus including the ground state),
we have
\BE
\sum_{A^*} \sigma_{f A^*} = \sigma_{f},
\EE
that is, the same as the cross section of the $\GG\rightarrow
f$ process alone, without any higher order processes, as given above.

One can make an estimate of the importance of the higher order
processes in the following way: The luminosity especially at high
invariant masses peaks at $b \approx b_{min} = 2 R$ and one can
therefore make the approximation
\BE
\sigma_{f,A} \approx P_{A}(2R)  
\sigma_{\GG\rightarrow f},
\EE
for the cross section $\sigma_{f,A}$, that is $\GG\rightarrow
f$ with the nucleus in the ground state in the final state and
\BE
\sigma_{f,A^*} \approx \left(1-P_{A}(2R)\right)
\sigma_{\GG\rightarrow f}
\EE
for the nucleus to be excited.
The impact parameter dependent probability for the excitation of the
nucleus can be found in
\cite{BertulaniB88,BaronB94,VidovicGS93,BaltzRW96}. They are strongly
dependent on $A$.
Especially important (see also Sec.~\ref{sec_ga}) is the giant dipole
resonance (GDR), a highly coherent excited state, which can be
interpreted as the movement of all neutrons against all protons. The
probability of such an excitation of a nucleus with charge $Z_2$ and 
nucleon number
$A_2$ ($N_2=A_2-Z_2$) by a nucleus with charge number $Z_1$ is given
in \cite{BertulaniB88} as
\BE
P_{GDR}(b) = 1-\exp(-S/b^2) ,
\EE
with 
\BE
S=5.45 \times 10^{-5} \frac{Z_2^2N_2 Z_2}{A_2^{2/3}} fm^2,
\label{eq_trks}
\EE
where the Thomas-Reiche-Kuhn (TRK) sum rule was used and the
position of the GDR is given by $80 A^{-1/3}$~MeV.
For Pb-Pb we get a value of $S=(10.4 fm)^2$. A parameterization which
includes also all additional contributions (quasideuteron absorption,
nucleon excitation, etc.) was given in \cite{BaltzS98}; they get a
value of $S=(17.4 fm)^2$. For the system Ca-Ca we get $S=(0.86 fm)^2$
using Eq.~(\ref{eq_trks}). 

From this estimate one finds that
about 75\% of the photon-photon events in Pb-Pb collisions
are accompanied by an excitation, about
40\% leading to the GDR. For Ca-Ca these are less than 2\% of all
events. As each of the ions can be excited, we get a total probability
for excitation of at least one of them of about 2\% for Ca-Ca and of
about 95\% for Pb-Pb (65\% from GDR). These effects are therefore
dominant for Pb-Pb, but almost unimportant for Ca-Ca.

\begin{figure}[tbhp]
\begin{center}
\ForceHeight{7cm}
\BoxedEPSF{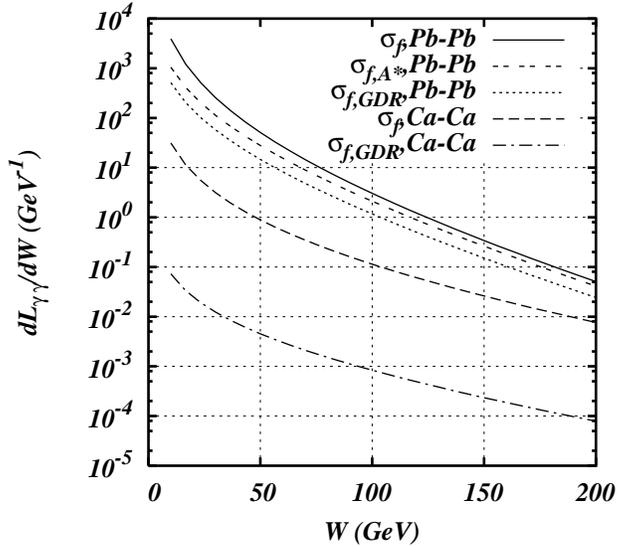}
\end{center}
\caption{\it
For Pb-Pb collisions a substantial portion of the total
luminosity ($\sigma_f$, solid curve) comes from events with are accompanied by
the excitation of at least one ion ($\sigma_{A*}$, double-dashed
curve) The dominant contribution is to the giant dipole resonance
($\sigma_{GDR}$, dotted curve). For Ca-Ca 
collisions this is no longer very important. Shown are full
calculation for the LHC ($\G\approx2950$ for Pb-Pb, 3750 for Ca-Ca).} 
\label{fig_lumgdr}
\end{figure}

A detailed calculation of the cross section using the full $b$
dependence is shown in Fig.~\ref{fig_lumgdr}. It shows that the
excitation probability is lower than the estimate given above.
%
% Figure Luminosity with GDR
%

The GDR excitation is followed most of the time by neutron
evaporation. Also other photon induced reactions 
predominantly lead to the emission
of individual nucleons. This emission of relatively low energy
nucleons in the nucleus rest frame leads to high energetic neutrons
(protons) with energies of about 3 TeV (LHC) or 100 GeV (RHIC). The
neutrons can possibly be detected in a zero degree calorimeter.

\begin{figure}[tbhp]
\begin{center}
\ForceWidth{0.6\hsize}
\BoxedEPSF{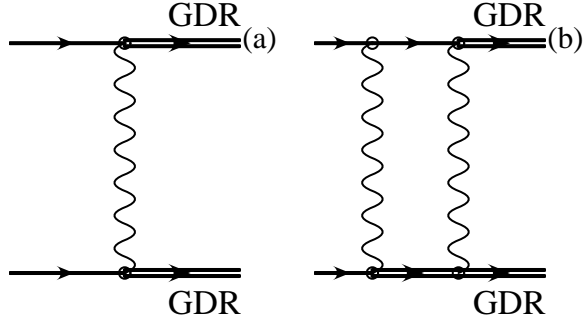}
\end{center}
\caption{\it
The mutual excitation process (a), where both ions are excited due to
one-photon exchange, becomes less important at larger $A$
compared to the second order process (b).
}
\label{fig_mutual}
\end{figure}

In \cite{Baltz98} it was proposed to use the mutual emission of
neutrons from both nuclei as a measure of the beam luminosity at RHIC.
Using the coincidence of two neutrons in the very forward and backward
direction other sources of neutrons can be suppressed
effectively. Since the photonuclear processes are large and well
understood they can lead to a good determination of the luminosity.
The authors of \cite{Baltz98} estimate to be able to determine the
luminosity to about 5\%. It seems interesting to note that the
$A$ dependence of the excitation cross section 
with two photons is given approximately
by $10^{-9} A^6 $~mb\cite{Baur90d}, the one-photon exchange cross
section is given approximately by $10^{-5} A^2$~mb
\cite{HenckenTB96}, see  Fig.~\ref{fig_mutual}. So for
nuclei heavier than about C, the two-photon mechanism is dominant over
the one-photon mechanism.
%
% figur mutual versus second order
%

\begin{figure}[tbhp]
\begin{center}
\ForceWidth{0.6\hsize}
\BoxedEPSF{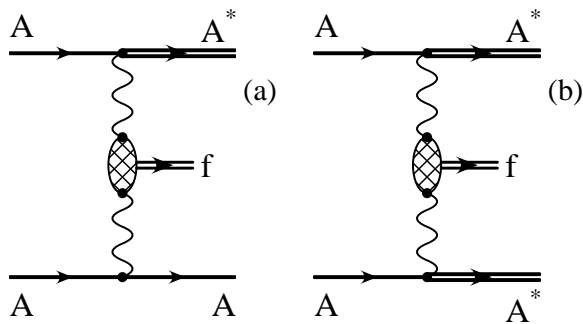}
\end{center}
\caption{\it
In photon-photon collisions, one (Fig.~(a)) and even both (Fig.~(b))
nuclear vertices can be inelastic, leading to excited nuclei.
}
\label{fig_iepa}
\end{figure}

In the calculation of the luminosity we were always assuming
that both nuclei remain in their ground state. But nuclei are weakly
bound composite systems and it is possible that the photon emission
leads to their excitation (see also Fig.~\ref{fig_iepa}). We
distinguish two different types: those leading to an excited
nucleus with a well defined excitation and the
incoherent photon emission from individual protons (or even quarks
within the protons), which are best treated as inclusive processes,
summing over all excitation energies.

An equation for the inelastic photon-emission to a discrete state, was
derived in \cite{HenckenTB95} (see also \cite{BudnevGM75})
using plane waves and therefore not
subtracting the central collisions.  It was applied to nuclear
excitation, as well as to the case of the proton-Delta transition
\cite{Conradt96}. The equivalent photon number can be expressed in
terms of the structure functions $C$ and $D$ of the general hadron
tensor 
\begin{eqnarray}
\overline{\sum_{M_i M_f}} \Gamma^{\mu*} \Gamma^\nu
&=&
\left[g^{\mu\nu}-\frac{q^\mu q^\nu}{q^2}\right] C + 
\left[P^\mu - \frac{qP}{q^2} q^\mu\right]
\left[P^\nu - \frac{qP}{q^2} q^\nu\right] D,
\end{eqnarray}
where $\Gamma_\mu$ denotes the nuclear four-current. One obtains
\BE
n(\omega)= \int \frac{(-2C+q_\perp/\omega^2 \ P^2 D)
\omega^2}{(2\pi)^3 2 E P (q^2)^2} d^2q_i,
\EE
where $E$ and $P$ are energy and momentum of the nucleus. Whereas in
the elastic case $q^2$ was given by $-(\omega^2/\G^2+q_\perp^2)$, it
is here 
$\approx -\left[ \frac{\omega}{\G} \left(\frac{\omega}{\G}+2 \Delta
\right)\right]$, where $\Delta$ is the excitation energy.
Using the Goldhaber-Teller model, the
transition to the GDR was found to be very small, below 1\% of the
elastic contribution. Using the elastic structure functions of the
proton in its usual dipole form (see, e.g., \cite{BudnevGM75}) the
elastic proton equivalent photon number was obtained analytically in
\cite{Kniehl91}. Quite similarly the equivalent photon number
corresponding to the $p-\Delta$ transition was obtained analytically
in \cite{Conradt96} using the structure functions of \cite{Chanfray93}
as
\begin{eqnarray}
n_{p\rightarrow\Delta}(\omega) &=& \frac{\alpha}{4\pi}
\frac{\mu^{*2}}{9 m^2} \left(\frac{m^*+m}{2m} \right)^2
\biggl[
t_{min} \biggl\{
\ln\left(\frac{t_{min}}{\Lambda^2+t_{min}}\right) + \frac{11}{6}
-\frac{2 t_{min}}{\Lambda^2+t_{min}} \nonumber\\ &&
+\frac{3 t_{min}^2}{2(\Lambda^2+t_{min})^2}
-\frac{t_{min}^3}{3(\Lambda^2+t_{min})^3}
\biggr\}
+\frac{\Lambda^8}{3(\Lambda^2+t_{min})^3}
\biggr] ,
\end{eqnarray}
with 
\BE
t_{min} = \frac{\omega^2}{\G^2} + \frac{(m^*-m)^2}{\G^2} + 2
\frac{\omega (m^*-m)}{\G}
\EE
and where $m^*=1232$ MeV is the mass of the $\Delta$, $m=938$ MeV 
the proton mass, $\Lambda^2=0.71$ GeV${}^2$ and
$\mu^*=9.42$. For not too large photon energy $\omega$ it is given by
a constant
\BE
n_{\Delta}(\omega) \approx \frac{\alpha}{4\pi} \frac{\mu^{*2}}{9
m^2}
\left(\frac{m^*+m}{2m} \right)^4 \frac{\Lambda^2}{3} .
\EE
This is an effect of the order of 10\% \cite{Conradt96}.
%
% figure inelastic emission
%

As we have seen above, the knowledge of impact parameter dependent
equivalent photon numbers is very helpful, as it allows in a direct
way to take the strong absorption into account (Eq.~(\ref{eq_fw1w2})).
The equivalent photon spectrum corresponding to a fast moving point
particle is given by Eq.~(\ref{eq_nomegab}). This result can now be
generalized to arbitrary charge-current distributions (also
quantum-mechanical transition currents in the framework of the Glauber
theory).

The equivalent photon spectrum corresponding to a spherically
symmetric charge distribution $\rho(r)$ moving with velocity $v$ at an
impact parameter $b$ is derived in \cite{BaurF91}. The Fourier
transformation of this charge distribution, with $\int d^3r \rho(r) =
Z$, is given by
\BE
f(k^2) = Z F(k^2) = \int d^3r \exp(-i\vec k \vec r) \rho(r).
\label{eq_zfk}
\EE

For $\G\gg1$ we only need the perpendicular component of the electric
and magnetic fields, $\vec E_\perp$ and $\vec B_\perp$. One finds
\cite{BaurF91}
\BE
\vec E_\perp =  \int \frac{d^2k_\perp}{(2 \pi)^2}
\frac{Z e}{v k^2} f(k^2) \exp(-i \vec k_\perp \vec b) \vec k_\perp ,
\label{eq_noemgaf}
\EE
where $k^2\approx - k_\perp^2 - (\omega/\G)^2$.
It can easily be seen, that $E_\perp
\parallel \vec b$ and one obtains the equivalent photon number as
\BE
N(\omega,b) = \frac{Z^2\alpha}{\pi^2} \left(\frac{c}{v}\right)^2
\frac{1}{b^2} 
\left| \int_0^\infty du u^2 J_1(u)
\frac{f(-\frac{x^2+u^2}{b^2})}{x^2+u^2}
\right|^2,
\label{eq_nwb}
\EE
where $x=\omega b /\G$.

In a similar way the equivalent photon spectrum of a point-like
magnetic dipole, moving with a constant velocity at a given impact
parameter, is calculated in \cite{Baron94}. This purely classical
result can also be interpreted quantum-mechanically by using the
corresponding electromagnetic matrix-elements.

An interesting question is the incoherent contributions due to the
protons inside the nucleus. We generalize
Eqs.~(\ref{eq_zfk})--(\ref{eq_nwb}) in the following way: The static
charge distribution is replaced by a transition charge density
$\rho_{f0}(r)$ and the Fourier transform of it is given by
\BE
f_{f0}(\vec k) = \int d^3r \exp(-i \vec k \vec r) \rho_{f0}(r) ,
\label{eq_ff0k}
\EE
where $f$ is some final state of the nucleus and 
\begin{eqnarray}
f_{f0}(\vec k) &=& \int d^3r \ d\xi \Psi_f^*(\xi) 
\sum_{i=1}^{Z} \delta(\vec r
- \vec r_i) \Psi_0(\xi) \exp(-i \vec k \vec r) \\
 &=& \sum_{i=1}^{Z} \int d\xi \Psi_f^*(\xi) \exp(-i \vec k \vec r_i)
\Psi_0(\xi) , 
\end{eqnarray}
where $\xi=r_1,r_2,\cdots,r_Z$ is the set of all proton coordinates
(for our argument we can neglect the neutrons). We get
the total incoherent contribution by summing over all states $f$
excluding the ground state $f=0$. The sum can be performed using the
closure relation and we obtain:
\begin{eqnarray}
S(\vec k,\vec k') &=& \sum_{f\not=0} f_{f0}^*(\vec k) f_{f0}(\vec k') 
\nonumber\\
& = & \int d\xi \ d\xi' \sum_{i,j} \Psi_0^*(\xi)\exp(i \vec k \vec
r_i) \Psi_f(\xi) \Psi_f^*(\xi')  \exp(-i \vec k' \vec {r_j'})
\Psi_0(\xi') 
\nonumber\\ &&
- f_{00}^*(\vec k) f_{00}(\vec k')\nonumber
\\
&=& \sum_{i,j} \int d\xi \left| \Psi_0(\xi) \right|^2 \exp(i\vec k
\vec {r_i} - i \vec k' \vec {r_j}) - f_{00}^*(\vec k) f_{00}(\vec k'). 
\end{eqnarray}
We split the sum over $i$,$j$ now into one for $i=j$ and one for
$i\not=j$, following \cite{deForestW66}. In the limit of no
correlation (not even Pauli correlations) we obtain
\BE
S(\vec k,\vec k') = Z F(\vec k-\vec k') - Z F(\vec k) F(\vec k') ,
\label{eq_skkp}
\EE
where we have introduced a normalized form factor $F(\vec k)$ of the
nucleus as $F(\vec k) = f_{00}(\vec k)/Z$, i.e., $F(0)=1$.

The equivalent photon number due to the incoherent contribution can
now be written as
\BE
N^{incoh}(\omega,b) = \frac{\alpha}{\pi^2}\left(\frac{c}{v}\right)^2
\frac{1}{b^2} \int \int d^2k_\perp d^2k_\perp' \vec k_\perp \vec
k_\perp' \frac{\exp(i \vec k_\perp \vec b - i \vec k_\perp' \vec
b)}{k^2 {k'}^2} S(\vec k,\vec k') ,
\label{eq_incoh1}
\EE
where $\vec k = (\vec k_\perp, \omega/v)$, $\vec k' = (\vec k_\perp',
\omega/v)$.  Using Eqs.~(\ref{eq_zfk}) and (\ref{eq_skkp})
and defining a thickness function
$T_z(b)$ --- see Eq.~(\ref{eq_thickness}) --- where the nucleon density
$n_A$ is now replaced by the charge density $\rho$), we have
\BE
N^{incoh}(\omega,b) = \int d^2r_\perp T_z(r_\perp)
N^{point}(\omega,\vec b +\vec r_\perp) - Z N^{form}(\omega,b) ,
\label{eq_nincohb}
\EE
where $N^{form}$ is the usual equivalent photon number for a given
form factor (divided by $Z^2$), see Eq.~(\ref{eq_nwb}), $N^{point}$
denotes the equivalent photon spectrum due to a point charge. For
$b=0$ this expression would diverge, and a suitable cut-off has to be
introduced ($b_{min}=R_{proton}$, or another value of $b_{min}$, for
which the equivalent photon approximation ceases to be valid).  If one
is integrating over all impact parameters (therefore also including
central collisions), the total incoherent equivalent photon number can
be defined:
\BE
n^{incoh}(\omega) = Z \left[ n^{point}(\omega) - n^{form}(\omega)
\right] ,
\label{eq_nincoh}
\EE
where $n^{i}(\omega)=\int d^2b N^{i}(\omega,b)$ and we have
used $\int d^2r_\perp T_z(\vec r_\perp)=Z$.

Incoherent scattering is a well known general phenomenon. E.g., the
blue sky is caused by the incoherent scattering of sunlight by gas
molecules or other randomly distributed dipole scatterer (see, e.g.,
p.418ff of \cite{JacksonED}). 
It can also be formulated in a way that makes a connection with the
parton model. As an example let us look at the contribution from
incoherent photon emissions of the quarks in the proton. 
In the parton model a proton in the
infinite momentum frame consists of partons (quarks, gluons,\dots).
We use the plane wave approach (see \cite{BudnevGM75}), where the
impact parameter dependence is not manifest. The equivalent photon
spectrum corresponding to a given proton final state
can be expressed in terms of the structure functions $W_1$ and
$W_2$. We only consider the dominant term corresponding to $W_2$. We
are also interested only in inclusive reactions, therefore we
integrate over the invariant mass of the final state $M^2$.  One has
\cite{BudnevGM75,HalzenM84}
\BE
n(\omega) = \frac{\alpha}{\pi} \int \int \frac{dQ^2 dM^2}{(Q^2)^2}
|q_\perp|^2 \frac{1}{2m} W_2(M^2,Q^2) ,
\EE
with $Q^2= Q_{min}^2(\omega) + q_\perp^2$.
Since $p=p'+q$ we obtain
\BE
M^2=m^2+2 m \nu - Q^2 ,
\EE
with $\nu=-pq/m$. (Since we assume that the photon is emitted, not
absorbed, from the nuclear system, our sign of $q$ and $\nu$ is
somewhat unconventional.) Introducing the scaling variable
$x=Q^2/(2m\nu)$, we can write in the scaling limit, see
e.g. \cite{HalzenM84}, 
\BE
W_2(M^2,Q^2) = \frac{1}{\nu} F_2(x) .
\EE
Changing the integration variable $dM^2$ to $dx$, we obtain
\BE
n(\omega) = \frac{\alpha}{\pi} \int dx F_2(x) \frac{1}{x} \int dQ^2
\frac{|q_\perp|^2}{(Q^2)^2} .
\label{eq_nwparton}
\EE

We express the structure function $F_2(x)$ in terms of the quark
distribution functions $f_{q_i|p}(x)$:
\BE
F_2(x) = x \sum_{q_i} e_i^2 f_{q_i|p}(x) ,
\EE
where $e_i$ is the charge of the quark $q_i$
($i=u,d,s,\cdots$). Eq.~(\ref{eq_nwparton}) now has an intuitive
interpretation: The proton consist of partons (=quarks) with an
momentum fraction $x$, they radiate as pointlike particles. This is
described by the $dQ^2$ integration. Taking also the energy loss of
the quarks into account in a more complete formula, the photon
spectrum from a quark is given by
\BE
f_{\G/q}(z) = \frac{\A}{2\pi}
\frac{1+(1-z)^2}{z} \log\left(\frac{Q_1^2}{Q_2^2} \right) ,
\EE
where $z$ is the ratio of photon-energy $\omega$ and energy of the
quark $x E$.

According to \cite{DreesGN94} we choose $Q_1^2$ to be the maximum
value of the momentum transfer given by $x_1 x_2 z_1 z_2 s/4 - m^2$
and the choice of the minimum $Q_2^2=1$~GeV$^2$ is made such that the
photons are sufficiently off shell for the quark-parton model to be
applicable. The inelastic contribution to the $\GG$-cross section is
then given by
\begin{eqnarray}
\sigma^{inel}_{pp} &=& \sum_{ij}  e_i^2 e_j^2 
\int^1_{\frac{4 m^2}{s}} dx_1
\int^1_{\frac{4 m^2}{s x_1}} dx_2
\int^1_{\frac{4 m^2}{s x_1 x_2}} dz_1
\int^1_{\frac{4 m^2}{s x_1 x_2 z_1}} dz_2 \nonumber\\
&& f_{q_i|p}(x_1,Q^2) f_{q_j|p}(x_1,Q^2)
f_{\G/q_i}(z_1) f_{\G/q_j}(z_2) \nonumber\\
&&\hat\sigma_{\GG}(W_{\GG} = \sqrt{x_1 x_2 z_1 z_2 s/4})
\end{eqnarray}
(Eq.~(3) of \cite{DreesGN94}) and where
$s$ is the invariant mass of the $pp$ system.
Of course the photons are now somewhat more off-shell than in the
elastic case, and for some cases it could be less safe to use the
assumption of real photons ($q^2=0$), when calculating the cross
section for the $\GG$-subprocess
$\hat\sigma$.
%
% inelastic processes
%
\begin{figure}[tbhp]
\begin{center}
\ForceHeight{6cm}
\BoxedEPSF{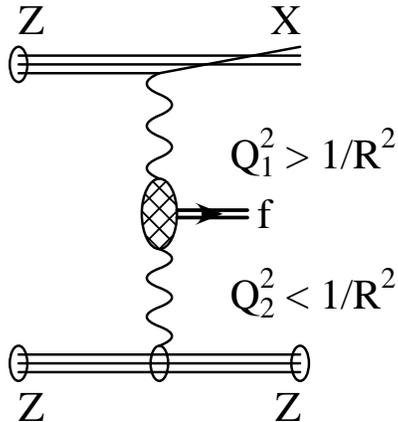}
\end{center}
\caption{\it With $Q^2 < 1/R^2$ the photon is emitted coherently from
all ``partons'' inside the ion. For $Q^2 \gg 1/R^2$ the ``partonic''
structure of the ion is resolved.
}
\label{fig_dis}
\end{figure}

A similar formula can also be written for the semielastic ( =
elastic-inelastic) cross section, see \cite{DreesGN94,OhnemusWZ94}.
$\GG$-luminosities are calculated according to this procedure in both
of these papers. The MRSD' parameterization \cite{Martin93a,Martin93b}
for the partonic densities is used in \cite{DreesGN94}, in
\cite{OhnemusWZ94} the simple parameterization $F_2= 0.16 \ln (1/x)$
was used. One finds that typically the inel-inel contribution is
largest, as the charges of the partons (quarks) are comparable to the
charges of the proton.

For the proton contribution to the photon spectrum of a heavy nucleus,
the situation is different. The coherent contribution is proportional
to $Z^2$, whereas they are only proportional to $Z$ for the
incoherent one. As $Z\gg1$ one expects the coherent part to be dominant.

%%%%%%%%%%%%%%%%%%%%%%%%%%%%%%%%%%%%%%%%%%%%%%%%%%%%%%%%%%%%%%%%%%%%%%
\section{$\G$-A interactions}
\label{sec_ga}

The cross section for the collisions of the equivalent photons of one 
nucleus with the other is given by (see Eq.~(\ref{eq_sigmac})):
\BE
\sigma = \int \frac{d\omega}{\omega} n(\omega) \sigma_{\G}(\omega).
\EE
where the equivalent photon number $n(\omega)$ is given in
Eq.~(\ref{eq_nomegaex}) and $\sigma_{\G}(\omega)$ is the photonuclear
cross section. This gives rise to many interesting phenomena ranging
from the excitation of discrete nuclear states, giant multipole
resonances (especially the giant dipole resonance), quasideuteron
absorption, nucleon resonance excitation to the nucleon continuum
(see, e.g.. \cite{PDG96,BauerSYP78}).  Photo-induced processes lead in
general to a change of the charge-to-mass ratio of the nuclei, and
with their large cross section they are therefore a serious source of
beam loss. Especially the cross section for the excitation of the the
giant dipole resonance, a collective mode of the nucleus, is rather
large for the heavy systems (of the order of 100b). For a recent
discussion see \cite{NorburyW98}.  The cross section scales
approximately with $Z^{10/3}$.  (Another serious source of beam loss,
the $\EPEM$ bound-free pair creation will be discussed in
Sec.~\ref{chap_epem}). The contribution of the nucleon resonances
(especially the $\Delta$ resonance) has also been confirmed
experimentally in fixed target experiments with 60 and~200 GeV/A
(heavy ions at CERN, ``electromagnetic spallation'')
\cite{BrechtmannH88a,BrechtmannH88b,PriceGW88,Pshenichnov98}. For
details of these aspects, we refer the reader to
\cite{KraussGS97,VidovicGS93,BaltzRW96,BaurB89}, where scaling laws,
as well as detailed calculations for individual cases are given.

Recently the total dissociation cross section for different ion
species was studied in an experiment at CERN \cite{Datz97}. There it
was found that this cross section is dominated at medium and large $Z$
by the electromagnetic dissociation, with the region of the GDR
contributing with about 80\%. The theoretical calculations, which can
be considered to be fairly reliable and detailed, see
e.g. \cite{NorburyB93}, agree quite well with the experiments apart
from an additional effect, which can be parameterized as
$\sigma_{add}=0.12 Z$~barn, a very sizeable effect (For large $Z$ it
is even larger than the nuclear cross section).  It is tempting to
guess that the $Z$ dependence is due to an incoherent effect of the
$Z$ protons in a nucleus. However the corresponding incoherent photon
number (Eq.~(\ref{eq_nincohb})) is very small for the relevant region
$b>R_1+R_2$. Therefore we exclude an incoherent effect as the
explanation of the anomaly observed in \cite{Datz97}.

The interaction of quasireal photons with protons has been studied
extensively at the electron-proton collider HERA (DESY, Hamburg), with
$\sqrt{s} = 300$~GeV ($E_e=27.5$~GeV and $E_p=820$~GeV in the
laboratory system). This is made possible by the large flux of
quasi-real photons from the electron (positron) beam (for a review see
\cite{Abbiendi97}). The obtained $\G p$ center-of-mass energies (up to
$W_{\G p}\approx200$~GeV) are an order of magnitude larger than those
reached by fixed target experiments. Similar and more detailed
studies will be possible
at the relativistic heavy ion colliders RHIC and LHC, due to the
larger flux of quasireal photons from one of the colliding nuclei. In
the photon-nucleon subsystem, one can reach invariant masses $W_{\G
N}$ up to $W_{\G N,max}=\sqrt{4 W_{max} E_N} \approx 0.8 \G
A^{-1/6}$~GeV. In the case of RHIC (${}^{197}$Au, $\G=106$) this is about
30~GeV, for LHC (${}^{208}$Pb, $\G=2950$) one obtains 950~GeV. Thus
one can study physics quite similar to the one at HERA, with nuclei
instead of protons. Photon-nucleon physics includes many aspects, like
the energy dependence of total cross-sections, diffractive and
non-diffractive processes (see, e.g., \cite{Abbiendi97}). An important
subject is elastic vector meson production $\G p \rightarrow V p$
(with $V=\rho,\omega,\phi,J/\Psi,\dots$). A review of exclusive
neutral vector meson production is given in \cite{Crittenden97}.  
The diffractive production of vector mesons allows one to get insight
into the interface between perturbative QCD and hadronic
physics. Elastic processes (i.e., the proton remains in the ground
state) have to be described within nonperturbative (and therefore
phenomenological) models. It
was shown in \cite{RyskinRML97} that diffractive (``elastic'') $J/\Psi$
photoproduction is a probe of the gluon density at $x\approx
\frac{M_{\Psi}^2}{W_{\G N}^2}$ (for quasireal photons).  Inelastic
$J/\Psi$ photoproduction was also studied recently at HERA
\cite{Breitweg97}.
Going to the hard exclusive photoproduction
of heavy mesons on the other hand, perturbative QCD is
applicable. Recent data from HERA on the photoproduction of $J/\Psi$
mesons have shown a rapid increase of the total cross section with
$W_{\G N}$, as predicted by perturbative QCD.
Such studies could be extended to photon-nucleus
interactions at RHIC, thus complementing the HERA studies. Equivalent
photon flux factors are large for the heavy ions due to coherence. On
the other hand, the A-A luminosities are quite low, as compared to
HERA. Of special interest is the coupling of the photon of one nucleus
to the Pomeron-field of the other nucleus. Such studies are envisaged
for RHIC, see \cite{KleinS97a,KleinS97b,KleinS95a,KleinS95b} where
also experimental feasibility studies were performed.

It is useful to have estimates of the order of magnitude of vector
meson production in photon-nucleon processes at RHIC and LHC. Let us
assume a cross-section that rises with the $\G p$ center of mass
energy approximately with a power law:
\BE
\sigma_{\G} = \sigma_0 \left(\frac{W_{\G N}}{W_0}\right)^{\beta} ,
\EE
with $W_0$ chosen to be 1~GeV and $\beta\approx 0.22$ for
$V=\rho,\omega,\phi$, and $\beta\approx 0.8$ for $V=J/\Psi$. Also
the total hadronic interaction cross section can be parameterized in
this form with $\beta\approx 0.16$. From Fig.~17 of \cite{Abbiendi97}
or Fig.~5 of \cite{Miller98} one has for $\sigma_0 \approx 50$~$\mu$b
for the total hadronic cross section, 5~$\mu$b for $V=\rho$,
$0.5$~$\mu$b for $V=\omega$, $\phi$ and $10^{-3}$~$\mu$b for
$V=J/\Psi$. Making use of the photon number of
Eq.~(\ref{eq_nomegaapprox}) the total cross section for vector meson
production on the reaction $Z+p \rightarrow Z+p+V$ due to the
equivalent photon spectrum of the nucleus $Z$ is obtained as
\begin{eqnarray}
\sigma &=& \frac{2 Z^2 \alpha}{\pi} \sigma_0
\left(\frac{2 m_N \G_p}{R W_0^2}\right)^{\beta/2} \times
\nonumber\\&&
\left[
\left(
1 - \left( \frac{\omega_{min} R}{\G_p} \right)^{\beta/2}
\right)
\frac{4}{\beta^2}
+
\left( \frac{\omega_{min} R}{\G_p} \right)^{\beta/2}
\frac{2 \ln \left(\omega_{min} R /\G_p\right)}{\beta}
\right] ,
\end{eqnarray}
where some minimum value of the energy of the equivalent photon is
used, say $\omega_{min}=1$~GeV for the total hadronic cross section, 
$\omega_{min}=2$~GeV for $V=\rho,\omega$ and~$\phi$ and
$\omega_{min}=10$~GeV for $V=J/\Psi$, and $\omega_{max}=\G_p/R$. 
For $\beta=0$ one obtains in a similar way
\BE
\sigma = \frac{Z^2 \alpha}{\pi} \sigma_0 \left[
\ln\left(\frac{\G_p}{\omega_{min}R}\right) \right]^2 .
\EE
The
Lorentz factor $\G_p$ of the nucleus $Z$, as viewed from the proton,
is given by
\BE
\G_p = 2 \G^2-1 .
\EE
We assume that a proton and a nucleus $Z$ collide, with the same value
$\G$ (see Fig.~\ref{fig_collision}, where one nucleus is replaced by a
proton). We obtain the following numbers, shown in Table~\ref{tab_vm}.
%
% Estimate of Vector meson production in Z+p
%
\begin{table}[tbhp]
\begin{center}
\begin{tabular}{|c|c|c|c|c|}
\hline
&\multicolumn{2}{c|}{RHIC}&\multicolumn{2}{c|}{LHC}\\
\hline
Ion & I & Au & Ca & Pb \\
$\G$ & 111 & 106 & 3750 & 2950\\
$R$(fm) & 6 & 7 & 4 & 7\\
\hline
$\sigma_{tot}$ (mb) & 20 & 40 & 15 & 200\\
$\sigma_\rho$ (mb) & 2 & 3.5 & 1.5 & 25\\
$\sigma_\omega$, $\sigma_\phi$ (mb) & 0.2& 0.35 & 0.15 & 2.5\\
$\sigma_{J/\Psi}$ ($\mu$b) & 0.5 & 1.5 & 3 & 40\\
\hline
\end{tabular}
\end{center}
\caption{\it
The expected cross sections for the elastic vector meson production on
a proton induced by the equivalent photons of a nucleus $Z$ are given
for RHIC and LHC conditions. Also shown are the total electromagnetic 
cross sections $\sigma_{tot}$. See text for details.}
\label{tab_vm}
\end{table}

The numbers in Table~\ref{tab_vm} refer to the photoproduction on one
proton. In $AA$ collisions there is incoherent photoproduction on the
individual $A$ nucleons. Shadowing effects will occur in the nuclear
environment and it will be interesting to study these. There is also
the coherent contribution where the nucleus remains in the ground
state. 
Due to the large momentum transfer, the total (angle integrated)
coherent scattering shows an undramatic $A^{4/3}$ dependence. This is
in contrast to, e.g., low energy $\nu$A elastic scattering, where the
coherence effect leads to an $A^2$ dependence, which is
relevant for the stellar collapse, see, e.g., \cite{FreedmanST77},
where also a pedagogical general discussion of coherence effects is
given. In addition there are inelastic contributions, where the
proton (nucleon) is transformed into some final state $X$ during the
interaction (see \cite{Breitweg97}). The experimental possibilities of
this at RHIC are investigated in \cite{KleinS97a,KleinS95a,KleinS95b}.

At the LHC one can extend these processes to much higher invariant
masses $W$, therefore much smaller values of $x$ will be probed.
Whereas the $J/\Psi$ production at HERA was measured up to invariant
masses of $W\approx 160$~GeV, the energies at the LHC allow for
studies up to $\approx 1$~TeV.

At the FELIX detector at LHC \cite{Felix97} hard diffractive vector
meson photoproduction can be investigated especially well in $AA$
collisions. In comparison to previous experiments, the very large
photon luminosity should allow observation of processes with quite
small $\G p$ cross sections, such as $\Upsilon$-production. For more
details see \cite{Felix97}.

\begin{figure}[tbhp]
\begin{center}
\ForceHeight{4cm}
\BoxedEPSF{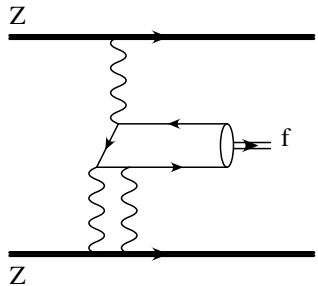}
\end{center}
\caption{\it
By using three (and more) photon processes also states that are
forbidden in the two-photon fusion process can be
produced. Interesting final states $f$ are, for example,
orthopositronium, orthomuonium, or the vector mesons
$\rho$,$\omega$,$\phi$. 
}
\label{fig_threephotons}
\end{figure}

$C=-1$ vector mesons can be produced in principle by the fusion of
three (or, less important, five, seven, \dots) equivalent photons (see
Fig.~\ref{fig_threephotons}). The cross section scales with
$Z^6$. This graph has been calculated by I. Ginzburg et
al. \cite{Ginzburg97} using the methods of
\cite{Ginzburg87,Ginzburg88,Ginzburg92}. It is smaller than the
contributions discussed above, even for nuclei with large $Z$. Quite
similarly the QED analogue of the $C=-1$ mesons, the ortho-states of
positronium, muonium, or tauonium can be calculated
\cite{Gevorkyan98}.
%
% Figure para- / ortho positronium
%

One might have thought that these three- and more photon contributions
are of the same order of magnitude, since $\ZA\le1$. But there is
another scale: As one can see from Eq.~(\ref{eq_pbapprox}) below,
there is another factor $1/(m b)$. For electrons the important
range $b \approx 1/m$ and this factor therefore is of the order of
unity. From this one expects that orthopositronium production can be
similar in size than parapositronium production. For heavier systems
the impact parameter range $b$ cannot be smaller than $R$, the nuclear
radius. Putting $1/R=\Lambda \approx 30$~MeV (for $R\approx 7fm$), the
scale factor is $\Lambda/M$ with $M$ the mass of the produced particle
$M=m_\mu$, \dots. This factor is always (much) smaller than 
one, thus the production via these higher order processes is small.
Of course the above arguments are rather qualitative and should be
complemented by more detailed calculations. Such calculations can be
done using the technique developed for vector meson production in
proton-proton collisions via gluon-exchange processes
\cite{Ginzburg87,Ginzburg88,Ginzburg92}.

As another possibility we mention photon-gluon fusion leading to the
production of $c\bar c$ and $b\bar b$ quark pairs. It was suggested in
Ref.~\cite{HofmannSSG91} as a possibility to deduce the in-medium
gluon distribution. Further studies were done in
Refs.~\cite{BaronB93,GreinerVHS95}, and this possibility is reviewed
in \cite{KraussGS97}.

%%%%%%%%%%%%%%%%%%%%%%%%%%%%%%%%%%%%%%%%%%%%%%%%%%%%%%%%%%%%%%%%%%%%%%
\section{Photon-Photon Physics at various invariant mass scales}
\label{chap_proc}

Scattering of light on light, while absent in classical Maxwell
electrodynamics, takes place due to quantum effects, like pair
creation. At low energies, photon-photon scattering is dominated by
electron intermediate states, the scattering of light on light occurs
in higher orders via an electron loop, see, e.g., \cite{Weinberg97}.
The lowest order process is $\EPEM$ pair creation
and is well described by QED.

Up to now photon-photon scattering has been mainly studied at $\EPEM$
colliders. Many reviews \cite{BudnevGM75,KolanoskiZ88,BergerW87}
as well as conference reports
\cite{Amiens80,SanDiego92,Sheffield95,Egmond97} exist. The
traditional range of invariant masses has been the region of mesons,
ranging from $\pi^0$ ($m_{\pi^0}=135$~MeV) up to about $\eta_c$
($m_{\eta_c}=2980$~MeV). Recently the total $\GG\rightarrow$~hadron
cross-section has been studied at LEP2 up to an invariant mass range
of about 70~GeV \cite{L3:97}. We are concerned here mainly with the
invariant mass region relevant for RHIC and LHC (see the
$\GG$-luminosity figures below). Apart from the production of $\EPEM$
(and $\MUPM$) pairs, the photons can always be considered as
quasireal. The cross section section for
virtual photons deviates from the one for real photons only for $Q^2$,
which are much larger then the coherence limit $Q^2\lesssim 1/R^2$
(see also the discussion in \cite{BudnevGM75}). For real photons
general symmetry requirements restrict the possible final states, as
is well known from the Landau-Yang theorem \cite{Yang48}. Especially
it is impossible to produce spin 1 final states. In $\EPEM$
annihilation only states with $J^{PC}=1^{--}$ can be produced
directly. Two photon collisions give access to most of the $C=+1$
mesons.

\begin{figure}[tbhp]
\begin{center}
\ForceWidth{0.9\hsize}
\BoxedEPSF{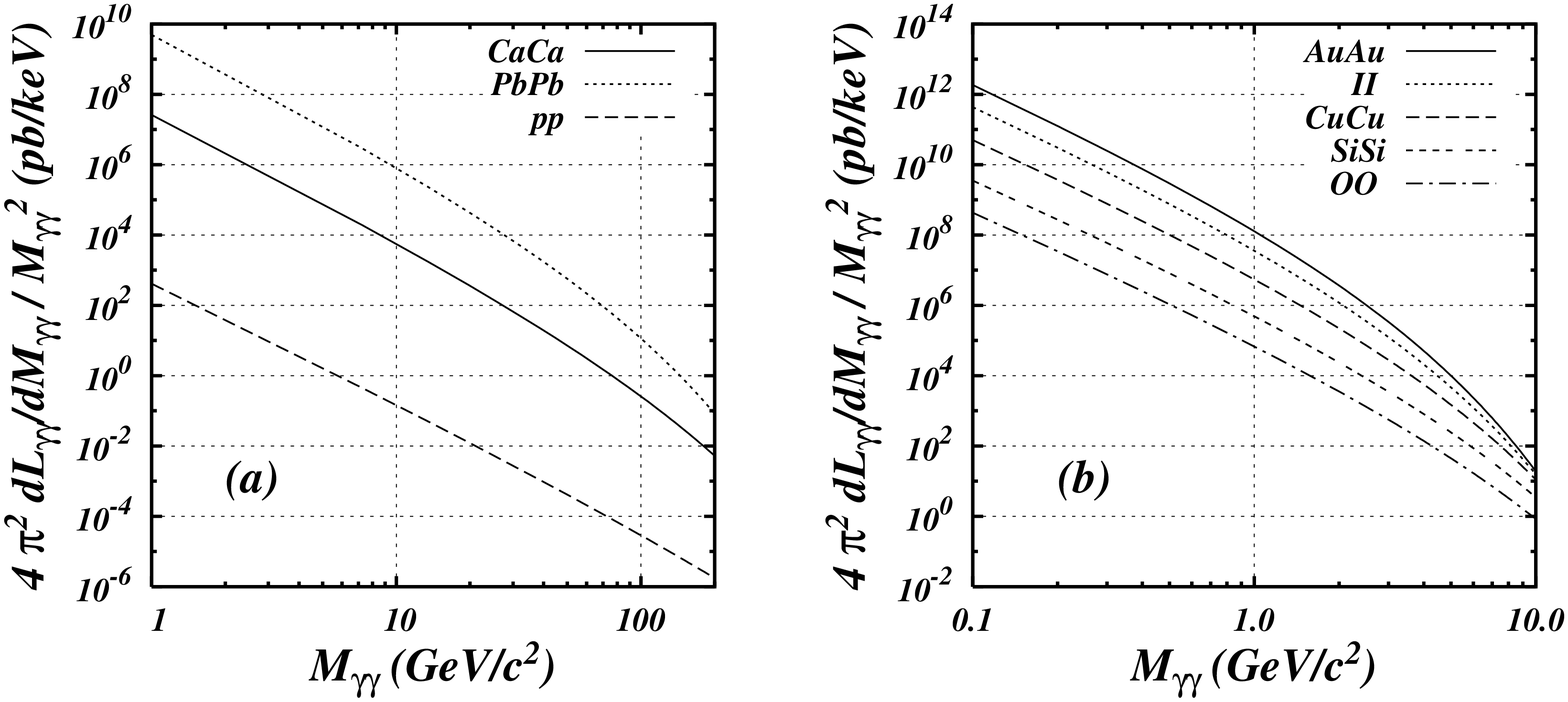}
\end{center}
\caption{\it
The universal function $4\pi^2 dL_{\GG}/dM_{\GG} /M_{\GG}^2$ is
plotted for different ion species at LHC (a) and RHIC (b). For the
parameters used see table~\protect\ref{tab_lum}.
}
\label{fig_sigmagamma}
\end{figure}

The cross section for $\GG$-production in a heavy ion collision
factorizes into a $\GG$-luminosity function and a cross-section
$\sigma_{\GG}(W_{\GG})$ for the reaction of the (quasi)real photons
$\GG \rightarrow f$, where $f$ is any final state of interest (see
Eq.~(\ref{eq_sigmaAA}).  When the final state is a narrow resonance,
the cross-section for its production in two-photon collisions is given
by
\BE
\sigma_{\GG\rightarrow R}(M^2) =
 8 \pi^2 (2 J_R+1) \Gamma_{\GG}(R) \delta(M^2-M_R^2)/M_R ,
\label{eq_nres}
\EE
where $J_R$, $M_R$ and $\Gamma_{\GG}(R)$ are the spin, mass and
two-photon width of the resonance $R$. This makes it easy to calculate
the production cross-section $\sigma_{AA\rightarrow AA+R}$ of a
particle in terms of its basic properties. In
Fig.~\ref{fig_sigmagamma} the function $4\pi^2 dL_{\GG}/dM /M^2$ is
plotted for various systems. It can be directly used to calculate the
cross-section for the production of a resonance $R$ with the formula
\BE
\sigma_{AA\rightarrow AA+R}(M) = (2 J_R +1) \Gamma_{\GG} \frac{4 \pi^2
dL_{\GG}/dM}{M^2} .
\label{eq_aar}
\EE
We will now give a general discussion of possible photon-photon
physics at relativistic heavy ion colliders. Invariant masses up to
several GeV can be reached at RHIC and up to about 100 GeV at LHC.
%
% Figure Luminosity function
%

We can divide our discussion into the following two main subsections:
Basic QCD phenomena in $\GG$-collisions (covering the range of meson,
meson-pair production, etc.) and $\GG$-collisions as a tool for new
physics, especially at very high invariant masses.

\subsection{Basic QCD phenomena in $\GG$-collisions}

\subsubsection{Hadron spectroscopy: Light quark spectroscopy}

One may say that photon-photon collisions provide an independent view
of the meson and baryon spectroscopy. They provide powerful
information on both the flavor and spin/angular momentum internal
structure of the mesons. Much has already been done at
$\EPEM$ colliders. For a 
review see, e.g., \cite{Cooper88}. Light quark spectroscopy is very
well possible at RHIC, benefiting from the high
$\GG$-luminosities. Detailed feasibility studies exist
\cite{KleinS97a,KleinS97b,KleinS95a,KleinS95b}.  In this study, $\GG$
signals and backgrounds from grazing nuclear and beam gas collisions
were simulated with both the FRITIOF and VENUS Monte Carlo codes. The
narrow $p_\perp$-spectra of the $\GG$-signals provide a good
discrimination against the background. The possibilities of the LHC
are given in the FELIX LoI \cite{Felix97}.

The absence of meson production via $\GG$-fusion is also of great
interest for glueball search. The two-photon width of a resonance is a
probe of the charge of its constituents, so the magnitude of the
two-photon coupling can serve to distinguish quark dominated
resonances from glue-dominated resonances (``glueballs'').  In
$\GG$-collisions, a glueball can only be produced via the annihilation
of a $q\bar q$ pair into a pair of gluons, whereas a normal $q\bar
q$-meson can be produced directly, so we estimate
\BE
\frac{\sigma(\GG \rightarrow M)}{\sigma(\GG \rightarrow G)}
=
\frac{\Gamma(M \rightarrow \GG)}{\Gamma(G \rightarrow \GG)}
\sim
\frac{1}{\alpha_s^2} ,
\EE
where $\alpha_s$ is the strong interaction coupling constant and where
 $G$ is a ``glueball'', $M$ a normal $q\bar q$-meson.  Glueballs are
 produced most easily in gluon-rich environment. This is, e.g., the
 case in radiative $J/\Psi$ decays, $J/\Psi \rightarrow \G g g$.

In order to form a meson out of the gluon pair, they must first
annihilate into a $q\bar q$ pair. So we estimate
\BE
\frac{\Gamma(J/\Psi \rightarrow \G G)}{\Gamma(J/\Psi \rightarrow \G
M)} \sim \frac{1}{\alpha_s^2} .
\EE
The ``stickiness'' of a mesonic state is defined as
(see, e.g., \cite{Cartwright98})
\BE
S_X = \frac{\Gamma(J/\Psi \rightarrow \G X)}{\Gamma(J/\Psi \rightarrow
\G \G)} .
\EE
We expect the stickiness of all mesons to be comparable, while for
glueballs it should be enhanced by a factor of about $S_G / S_M
\approx 1/\alpha_s^4 \sim 20$,

In a recent reference \cite{Godang97} results of the search for $f_J
(2220)$ production in two-photon interactions were presented. There a
very small upper limit for the product of $\Gamma_{\GG} B_{K_sK_s}$
was given, where $B_{K_s K_s}$ denotes the branching fraction of
its decay into $K_s K_s$.  From this it was concluded that this is a
strong evidence that the $f_J(2220)$ is a glueball.

%%%%%%%%%%%%%%%%%%%%%%%%%%%%%%%%%%%%%%%%%%%%%%%%%%%%%%%%%%%%%%%%%%%%%%
\subsubsection{Heavy Quark Spectroscopy}

For charmonium production, the two-photon width $\Gamma_{\GG}$ of
$\eta_c$ (2960 MeV, $J^{PC} = 0^{-+}$) is known from experiment. But
the two-photon widths of $P$-wave charmonium states have been measured
with only modest accuracy.  For RHIC the study of $\eta_c$ is a real
challenge \cite{KleinS97b}; the luminosities are falling and the
branching ratios to experimental interesting channels are small.

In Table~\ref{tab_ggmeson} (adapted from table~2.6 of \cite{Felix97})
the two-photon production cross-sections
for $c\bar c$ and $b \bar b$ mesons in the rapidity range $|Y|<7$ are
given. Also given are the number of events in a $10^6$ sec run with
the ion luminosities of $4\times 10^{30}$cm${}^{-2}$s${}^{-1}$ for
Ca-Ca and $10^{26}$cm${}^{-2}$s${}^{-1}$ for Pb-Pb. Millions of
$C$-even charmonium states will be produced in coherent two-photon
processes during a standard $10^6$~sec heavy ion run at the LHC. The
detection efficiency of charmonium events has been estimated as 5\%
for the forward-backward FELIX geometry \cite{Felix97}, i.e., one can
expect detection of about $5\times 10^3$ charmonium events in Pb-Pb
and about $10^6$ events in Ca-Ca collisions. This is two to three
orders of magnitude higher than what is expected during five years of
LEP200 operation. Further details, also on experimental cuts,
backgrounds and the possibilities for the study of $C$-even bottonium
states are given in \cite{Felix97}.
%
% Meson production table
%
\begin{table}[hbt]
\begin{center}
\begin{tabular}{|l|c|c|r|c|c|c|}
\hline
State   & Mass,     & $\Gamma_{\GG}$ &
              \multicolumn{2}{|c}{$\sigma (AA\to AA+X)$} &
              \multicolumn{2}{|c|}{Events for $10^6$~sec} \\ 
\cline{4-7}
  & MeV     &  keV  &      Pb-Pb  &                Ca-Ca  &
                           Pb-Pb  &                Ca-Ca  \\
\hline
~~~$\eta'$  & 958 & 4.2 & 22 mb  & 125 $\mu$b  & $2.2 \times 10^7$
                                       & $5.0 \times 10^8$ \\
~~~$\eta_c$ & 2981 & 7.5 & 590 $\mu$b &3.8 $\mu$b &$5.9\times10^5$
                                       & $1.5 \times 10^7$ \\
~~~$\chi_{0c}$& 3415& 3.3& 160 $\mu$b &1.0 $\mu$b &$1.6\times10^5$
                                       & $4.0 \times 10^6$ \\
~~~$\chi_{2c}$& 3556 & 0.8 &160 $\mu$b& 1.0 $\mu$b&$1.6\times10^5$
                                       & $4.0 \times 10^6$ \\
~~~$\eta_b$   & 9366 & 0.43 &370 nb & 3.0 nb   & $370     $
                                              & $12000   $ \\
~~~$\eta_{0b}$& 9860 & $2.5\times10^{-2}$ & 18 nb & 0.14 nb & $18$
                                              & $640     $ \\
~~~$\eta_{2b}$& 9913 & $6.7\times10^{-3}$ & 23 nb & 0.19 nb & $23$
                                              & $76      $ \\
\hline
\end{tabular}
\end{center}
\caption{\it
Production cross sections and event numbers for heavy quarkonia
produced in a $10^6$ sec run in Pb-Pb and Ca-Ca collisions at the LHC
with luminosities
10$^{27}$ and 4$\times 10^{30}~{\rm cm}^{-2} {\rm sec}^{-1}$.
Adapted from \protect\cite{Felix97}.}
\label{tab_ggmeson}
\end{table}

\subsubsection{Vector-meson pair production. Total hadronic
cross-section} 

\begin{figure}[tbhp]
\begin{center}
\ForceWidth{0.6\hsize}
\BoxedEPSF{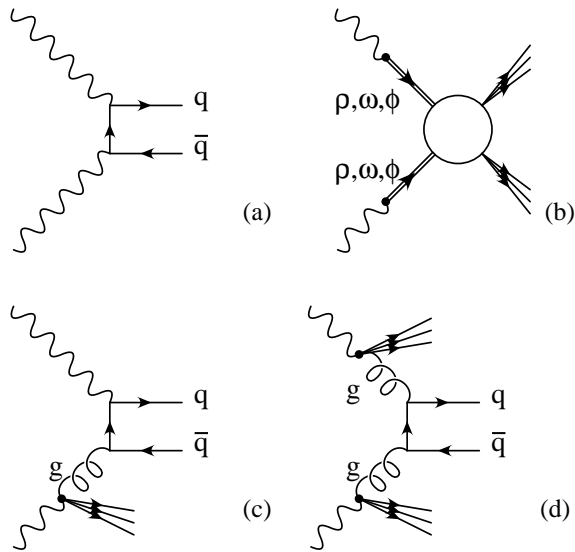}
\end{center}

\caption{\it
Diagrams showing the contribution to the $\GG\rightarrow$hadron
reaction: direct mechanism (a), vector meson dominance (b), single (c)
and double (d) resolved photons.
}
\label{fig_resolved}
\end{figure}

There are various mechanisms to produce hadrons in photon-photon
collisions. Photons can interact as point particles which produce
quark-antiquark pairs (jets) (see Fig.~\ref{fig_resolved}a), which
subsequently hadronize. Often a quantum fluctuation transforms the
photon into a vector meson ($\rho$,$\omega$,$\phi$, \dots) (VMD
component) opening up all the possibilities of hadronic
interactions (Fig.~\ref{fig_resolved}b). 
In hard scattering, the structure of the photon can be
resolved into quarks and gluons. Leaving a spectator jet, the quarks
and gluon contained in the photon will take part in the interaction,
some examples are given in Figs.~\ref{fig_resolved}c and d.  It is of
great interest to study the relative amounts of these components and
their properties.
%
% the resolve photon a and b
%

The L3 collaboration recently made a measurement of the total hadron
cross-section for photon-photon collisions in the interval $5 GeV <
W_{\GG} < 75 GeV$ \cite{L3:97}. It was found that the $\GG
\rightarrow$hadrons cross-section is consistent with the universal
Regge behavior of total hadronic cross-sections. They show a steep
decrease in the region of low center of mass energy followed by a slow
rise at high energies. It is parameterized as follows (see Eq.~(5) of
\cite{L3:97})
\BE
\sigma_{tot} = A (s/s_0)^\epsilon + B (s/s_0)^{-\eta} ,
\EE
with $\epsilon=0.0790\pm0.0011$, $\eta=0.4678\pm0.0059$,
$A=173\pm7$~nb, $B=519\pm 125$~nb, and $s_0=1$~GeV${}^2$. Using this
together with the effective luminosities (see Table~\ref{tab_lum}),
one expects about $3\times 10^6$ events/GeV at an invariant mass of
25~GeV and still $4\times 10^4$ events/GeV at $W_{\GG}=100$~GeV.

The production of vector meson pairs can well be studied at RHIC with
high statistics in the GeV region \cite{KleinS97a}.  For the
possibilities at LHC, we refer the reader to \cite{Felix97} and 
\cite{BaurHTS98}, where also experimental details and simulations are
described.

\subsection{$\GG$-collisions as a tool for new physics}

The high flux of photons at relativistic heavy ion colliders offers
possibilities for the search of new physics. This includes the
discovery of the Higgs-boson in the $\GG$-production channel or new
physics beyond the standard model, like supersymmetry or
compositeness.

Let us mention here the plans to build an $\EPEM$ linear collider.
Such future linear colliders will be used for $\EPEM$, $e\G$
and $\GG$-collisions (PLC, photon linear collider). 
The photons will be obtained by scattering of
laser photons (of eV energy) on high energy electrons ($\approx$ TeV
region) (see \cite{Telnov95}). Such photons in the TeV energy range
will be monochromatic and polarized. The physics program at such
future machines is discussed in \cite{ginzburg95}, it includes Higgs
boson and gauge boson physics and the discovery of new particles.

While the $\GG$ invariant masses which will be reached at RHIC will
mainly be useful to explore QCD at lower energies, the $\GG$ invariant
mass range at LHC --- up to about 100 GeV --- will open up new
possibilities.

A number of calculations have been made for a medium heavy standard
model Higgs \cite{DreesEZ89,MuellerS90,Papageorgiu95,Norbury90}. For
masses $m_H < 2 m_{W^\pm}$ the Higgs bosons decays dominantly into
$b\bar b$, whereas a heavier Higgs decays into a $W^+W^-$ pair. For
the $\GG\rightarrow H$ cross section we can use Eq.~(\ref{eq_aar}),
where the two-photon width of the Higgs bosons in the standard model
can be found, e.g., in \cite{DreesEZ89}. The calculations, using the
integrated luminosity of Table~\ref{tab_lum}, show that for Ca-Ca
collisions only about one Higgs boson is produced during one year of
the LHC operation. Therefore chances of finding the standard model
Higgs in this case are marginal \cite{BaurHTS98}.

An alternative scenario with a light Higgs boson was, e.g., given in
\cite{ChoudhuryK97} in the framework of the ``general two Higgs
doublet model''. Such a model allows for a very light particle in the
few GeV region. With a mass of 10~GeV, the $\GG$-width is about 0.1
keV (see Fig.~1 of\cite{ChoudhuryK97}). We get $2\times 10^3$ events
for Ca-Ca collisions, $40$ for $pp$ and 8 for Pb-Pb, with the
integrated luminosities of table~\ref{tab_lum}.  The authors of
\cite{ChoudhuryK97} proposed to look for such a light neutral Higgs
boson at the proposed low energy $\GG$-collider. We want to point out
that the LHC Ca-Ca heavy ion mode would also be very suitable for such
a search.

One can also speculate about new particles with strong coupling to the
$\GG$-channel. Large $\Gamma_{\GG}$-widths will directly lead to large
$\GG$ production cross-sections, see Eq.~(\ref{eq_aar}). We quote the
papers \cite{Renard83,BaurFF84}. Since the $\GG$-width of a resonance
is mainly proportional to the wave function at the origin, huge values
can be obtained for very tightly bound systems. Composite scalar
bosons at $W_{\GG}\approx 50$~GeV are expected to have $\GG$-widths of
several MeV \cite{Renard83,BaurFF84}. The search for such kind of
resonances in the $\GG$-production channel will be possible at
LHC. Production cross-section can be directly read off from
Eq.~(\ref{eq_aar}) and Fig.~\ref{fig_sigmagamma}. E.g., take $W_{\GG}=
50$~GeV and assume a width of $\Gamma_{\GG}=1 $~MeV, one obtains for a
scalar particle ($J_R=0$) $\sigma_{CaCa} \approx 1 MeV
10$~pb~keV${}^{-1} = 10$~nb. With an integrated luminosity of
4~pb${}^{-1}$ in the Ca-Ca mode, one obtains $4\times 10^4$ events.

In Refs. \cite{DreesGN94,OhnemusWZ94} $\GG$-processes at $pp$ colliders
(LHC) are studied. It is observed there that non-strongly interacting
supersymmetric particles (sleptons, charginos, neutralinos, and
charged Higgs bosons) are difficult to detect at the LHC. The
Drell-Yan and gg-fusion mechanisms yield low production rates for such
particles. Therefore the possibility of producing such particles in
$\GG$ interactions at hadron colliders is examined. Since photons can
be emitted from protons which do not break up in the radiation process
(see also Sec.~\ref{sec_lum}) clean events can be generated which
should compensate for the small number. Formula and graphs for the
production of supersymmetric particles are also given in
\cite{KraussGS97}, where also further references can be found.

In
\cite{DreesGN94} it was pointed out that at the high luminosity of
$L=10^{34}$cm${}^{-2}$s${}^{-1}$ at the LHC($pp$), one expects about
16 minimum bias events per bunch crossing. Even the elastic $\GG$
events will therefore not be free of hadronic debris. Clean elastic
events will be detectable at luminosities below
$10^{33}$cm${}^{-2}$s${}^{-1}$. This danger of ``overlapping events''
has also to be checked for the heavy ion runs, but it will be much
reduced due to the lower luminosities.

Detailed calculations have also been made for the production of a
charged chargino pair via $\GG\rightarrow \tilde\chi_1^+
\tilde\chi_1^-$. The production of these charginos can be studied via
their decay into a neutralino and a fermion-antifermion pair.
$\tilde\chi_1^\pm \rightarrow \tilde\chi_1^0 f_i \bar f_j$. The most
clean channel is into muons or electrons. Such an event would
therefore be characterized by two fermions of opposite charge
($e^+e^-$, $\mu^+\mu^-$ or $e^\pm\mu^\mp$) together with an unbalanced
transverse momentum. In order to be able to detect the missing
momentum, a closed geometry is needed.
Studies were made for this process as a function of the mass of the
chargino.
In this case the main background --- the production of a
$W^+ W^-$ pair also decaying into two leptons of opposite charge ---
was studied also. The cross section for this process was found to be
$3.6$~pb, comparable to the chargino pair production. But the harder
momentum distribution of this background process can be used to
distinguish it from the chargino production. Similar calculations have
also been made for $pp$ collisions \cite{OhnemusWZ94}.

%%%%%%%%%%%%%%%%%%%%%%%%%%%%%%%%%%%%%%%%%%%%%%%%%%%%%%%%%%%%%%%%%%%%%%
\section{Diffractive processes as background}
\label{sec_diffraction}

Diffractive processes are an important class of background to $\GG$
final states. As the nuclei can remain intact in these collisions, they
have the same signature as the photon-photon events. Therefore they
cannot be distinguished from each other. Diffractive events have
been studied extensively at HERA for photon-proton collisions. A
future program will also study diffractive processes involving nuclei
\cite{herafuture96}. Diffraction processes in $pp$ and $p\bar p$ are
also well known from studies at the Tevatron and ISR (CERN).

Diffractive events at high energies are best described within Regge
theory and in the language of the Pomeron (see,
e.g., \cite{Mueller98}). It is needless to say that
the possibility to study Photon-Pomeron and also Pomeron-Pomeron
collisions in peripheral collisions are interesting fields in
themselves. Especially photon-Pomeron fusion processes could be of
interest, as they allow for final states, which are not directly
possible in photon-photon events (see also the discussion about the
diffractive vector-meson production in $\G A$-collisions in
Sec.~\ref{sec_ga}). Here we restrict ourselves to the
estimate, how big the contribution of diffractive processes are
compared to photon-photon events. Of course it is difficult to give
quantitative answers at present.

A number of calculations were made within the phenomenological Dual
Parton Model (DPM) \cite{EngelRR97}. These calculations
\cite{EngelRR97} have been interpreted, that Pomeron-Pomeron fusion
dominates over the photon-photon cross section for almost all ions
used. Only for the very heavy ions, like Pb Pb, the photon-photon
process becomes comparable. Unfortunately these calculations were made
without the constraint that the nuclei remain intact in the final state.
As the nuclei are
only weakly bound system and the nuclear interaction strong, it is
very likely that a short range interaction between them leads to the
breakup of the nucleus. More refined calculations have been made in
the meantime \cite{Engel98}. The cross sections for diffractive
processes are then reduced roughly by two orders of magnitude for
Ca-Ca and by three orders of magnitude for Pb-Pb (both at LHC
conditions). Only for proton-proton collisions diffractive processes
can be expected to dominate over photon-photon ones.
Particle production from diffractive processes are also studied in
\cite{MuellerS91,SchrammR97}. They also find that the increase of the
cross section with mass number $A$ is much smaller than that for
electromagnetic processes. 

The problem of separating two-photon signals from background has been
studied in detail for RHIC conditions in \cite{NystrandK97}. Four
sources of background have been considered: Peripheral (hadronic)
nucleus-nucleus collisions, beam-gas interactions, gamma-nucleus
interactions and cosmic rays. In order to separate signals from
background, cuts have been developed which utilize the characteristics
of two-photon interactions; the most important of these cuts are
multiplicity and transverse momentum. It was shown in this reference,
to which we
refer the reader for details, that there are high rates of
$\GG$-interactions and that the signals can well be separated from
background. 

%%%%%%%%%%%%%%%%%%%%%%%%%%%%%%%%%%%%%%%%%%%%%%%%%%%%%%%%%%%%%%%%%%%%%%
\section{Electron-Positron Pair production and QED of strong fields}
\label{chap_epem}

Electrons (positrons) and to some extent also muons have a special
status, which is due to their small mass. They are therefore produced
more easily than other heavier particles and in the case of $\EPEM$
pair production lead to new phenomena, like multiple pair
production. In addition the Compton wave length of the electron
($\approx$ 386 fm) is much larger than the size of the nuclei
($\gtrsim$ 7 fm). This also means that the virtuality $Q^2$ of the
photons, which ranges from 0 up to the order of $1/R^2$ can be much
larger than the electron mass $m_e^2$. Whereas in all cases discussed
above 
we could treat the photons as being quasireal and relate their cross
section to the photon cross section, this is no longer the case here
in general and therefore corrections to the EPA are needed.

The muon has a Compton wavelength of about 2 fm. This length is of the
same order as the nuclear radius. We therefore expect that the EPA
will give more reliable results. Both electrons and muons can be
produced not only as free particles but also into an atomic states
bound to one of the ions.

\subsection{Free pair production, Strong field effects and multiple
pair production} 

We consider the $\EPEM$ pair production in the collision of two nuclei
with charges $Z_1$ and $Z_2$ and relative velocity $v$. For
$v\rightarrow 0$ the electrons and positrons can adjust adiabatically
to the motion of the nuclei and, with sufficiently high charge $Z_1$
and $Z_2$ one can enter a supercritical regime, where $(Z_1+Z_2) \A
>1$. Such a situation has been studied extensively at GSI and later at
Argonne; we refer the reader to the (vast) literature on this subject,
see \cite{Griffin98} where further references are given. We here study
the opposite region, with $v\approx c$. A very useful and complete
reference for this field is \cite{EichlerM95}.

The special situation of the electron pairs can already be seen from
the formula for
the impact parameter dependent probability in lowest order. Using 
EPA one obtains~\cite{BertulaniB88}
\BE
P^{(1)}(b) \approx \frac{14}{9 \pi^2} \left(Z \A\right)^4
\frac{1}{m_e^2 b^2} \ln^2 \left( \frac{\G \delta}{2 m_e b}\right) ,
\label{eq_pbapprox}
\EE
where $\delta\approx 0.681$ and $\G=2\G_{cm}^2-1$ the Lorentz factor
in the target frame, one can see that at RHIC and LHC energies and for
impact parameters of the order of the Compton wave length $b\approx
1/m_e$, this probability exceeds one. It was first described in
\cite{Baur90} how unitarity can be restored by considering the
production of multiple pairs\footnote{It is interesting to
remark that the fact that pair production in ion collisions grows
beyond the unitarity limit was already observed by Heitler
\cite{heitler34}. Of course at that time available energies made this
only ``of academic interest''.}.

Multiple pair production was later studied by a number of authors
\cite{Baur90c,BestGS92,RhoadesBrownW91,HenckenTB95a} using different
approximations. A general feature found by all was the fact that the
probability is given to a good approximation by a Poisson
distribution:
\BE
P(N,b) \approx \frac{\left[P^{(1)}(b)\right]^N}{N!} 
\exp\left[-P^{(1)}(b)\right] ,
\EE
where $P^{(1)}(b)$ is the single pair creation probability from
perturbation theory, see, e.g., Eq.~(\ref{eq_pbapprox}). Deviations
from this Poisson form were studied in \cite{HenckenTB95a} and were
found to be small at high energies.

\begin{figure}[tbhp]
\begin{center}
\ForceWidth{0.9\hsize}
\BoxedEPSF{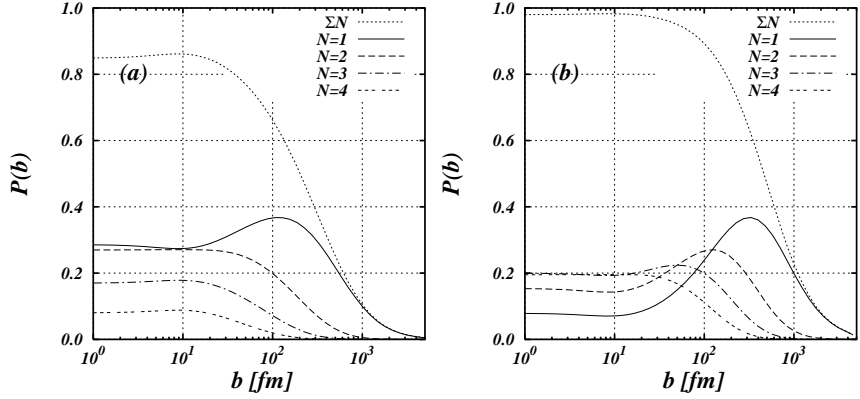}
\end{center}
\caption{\it
The impact parameter dependent probability to produce $N$
$\EPEM$-pairs ($N=1,2,3,4$) in one collision is shown for both RHIC
(a,$\G=106$,Au-Au) and LHC (b,$\G=2950$,Pb-Pb). Also shown is the
total probability to produce at least one $\EPEM$-pair. One sees that at
small impact parameters multiple pair production can be dominant over
single pair production.
}
\label{fig_pbee1}
\end{figure}

\begin{figure}[tbhp]
\begin{center}
\ForceHeight{7cm}
\BoxedEPSF{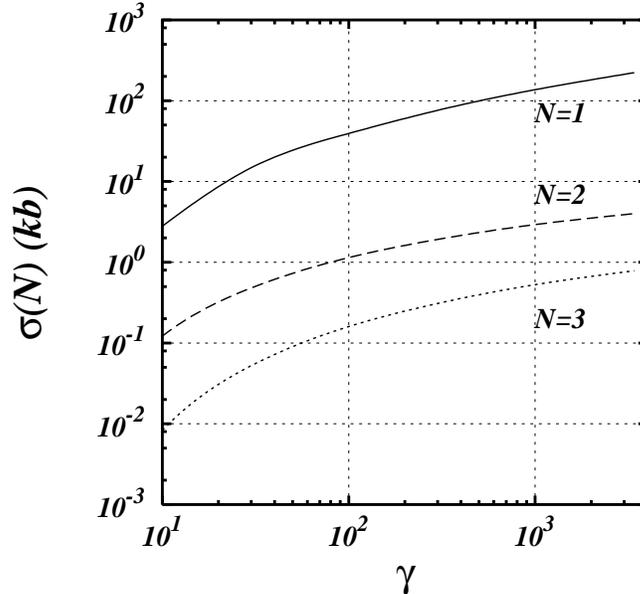}
\end{center}
\caption{\it
The total cross sections for the multiple pair production of up to
three pairs is shown as a function of the Lorentz factor $\G$. Shown
are the results for a Pb-Pb collision.
}
\label{fig_sigmaee}
\end{figure}

The impact parameter dependence of the lowest order process was
calculated in \cite{HenckenTB95b,Guclu95} (see also
Fig.~\ref{fig_pbee1}), the total cross section for the one-pair
production in \cite{Bottcher89}, for one and multiple pair production
in \cite{alscherHT97} (see Fig.~\ref{fig_sigmaee}).  Of course
the total cross section is dominated by the single pair production as
the main contribution to the cross section comes from very large
impact parameters $b$.  On the other hand one can see that for impact
parameters $b$ of about $2R$ the number of electron-positron pairs
produced in each ion collision is about 5 (2) for LHC with $Z=82$
(RHIC with $Z=79$). This means that each photon-photon event ---
especially those at a high invariant mass --- which occur
predominantly at impact parameters close to $b \gtrsim 2 R$ --- is
accompanied by the production of several (low-energy) $\EPEM$ pairs.
%
% Figuren
%
%
% Figuren sigma
%
%
% Figuren dsigma/dE, dsigma/dtheta
%
\begin{figure}[tbhp]
\begin{center}
\ForceWidth{0.9\hsize}
\BoxedEPSF{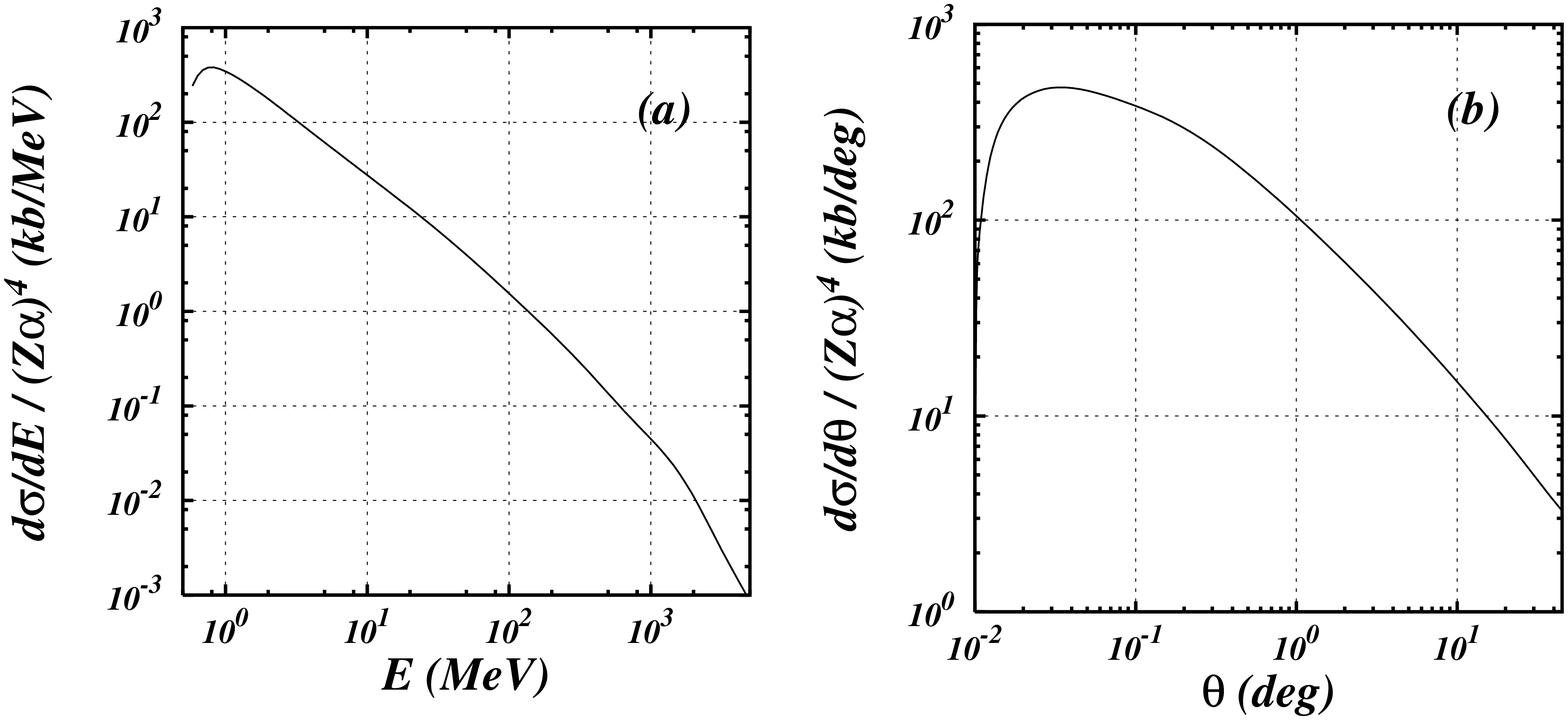}
\end{center}
\caption{\it
The single differential cross section for $\EPEM$ pair production at
LHC energies. Shown are the cross sections as a function of the energy
of either electron or positron (a), and as a function of the angle
with the beam axis $\theta$ (b).
}
\label{fig_eediff}
\end{figure}

As the total cross section for this process is huge (about 200~kb for
Pb at LHC, 30~kb for Au at RHIC), one has to take this process into
account as a possible background process. Most of the particles are
produced at low invariant masses (below 10 MeV) and into the very
forward direction (see Fig.~\ref{fig_eediff}). Therefore most of them
leave the detector along the beam pipe and are not observed. On the
other hand a substantial amount of them is left also at high energies,
e.g., above 1 GeV. These QED pairs also constitute a
potential hazard for detectors at the colliders.
In Table~\ref{tab_pairs1GeV} we show the cross
section for this process with the energy of either the electron or the
positron above a certain threshold. Singles angular
distributions of electrons (positrons) are calculated for peripheral
collisions using EPA in \cite{BaronB92}. Numerical results in
the relevant energy and angular ranges are presented there. The
physics is discussed in terms of easy to handle analytical formulae.
%
% table epem with threshold
%
\begin{table}[tbhp]
\begin{center}
\begin{tabular}{|c|rr|}
\hline
$E_{thr}$ (GeV) & $\sigma$(Pb-Pb,LHC) & $\sigma$(Ca-Ca,LHC)\\
\hline
0.25& 3.5  kb & 12  b\\
0.50& 1.5  kb & 5.5 b\\
1.0 & 0.5  kb & 1.8 b\\
2.5 & 0.08 kb & 0.3 b\\
5.0 & 0.03 kb & 0.1 b\\
\hline
\end{tabular}
\end{center}
\caption{\it
The cross sections of $e^+e^-$ pair production with {\em both} electron
and positron having an energy above a certain threshold value. Shown
are results for both LHC for two different ion species.}
\label{tab_pairs1GeV}
\end{table}

Differential production probabilities for $\GG$-dileptons in central 
relativistic heavy ion collisions are calculated using EPA and an
impact parameter formulation and compared to Drell-Yan and thermal
ones in \cite{Baur92,BaurB93b,Baur92b}. The very low $p_\perp$ values
and the angular distribution of the pairs give a handle for their
discrimination. Nuclear stopping leads to bremsstrahlung
pair-production and some modification of the $\GG$ dilepton
spectra. For details we refer the reader to these references. In
\cite{BaronB92b} the low energy dilepton spectrum in 16 GeV $\pi$-p
collisions was studied using the two-photon mechanism. It was found
that this mechanism could not explain the experimental data
\cite{stekas81}. This is in contrast to the findings of
\cite{BottcherSAE90}. 

In the Bethe-Heitler process $\G + Z \rightarrow \EPEM + Z$
higher-order effects are well known. Using Sommerfeld-Maue wave
functions higher order effects are taken into account. This leads to
a modification of the Born result. E.g., the total cross section (no
screening) for $\omega \gg m_e$ is given by (see \cite{LandauLQED})
\BE
\sigma = \frac{28}{9} Z^2 \A r_e^2 \left[ \ln \frac{2 \omega}{m_e} -
\frac{109}{42} - f(\ZA) \right] ,
\EE
with the higher-order term given by
\BE
f(\ZA) = (\ZA)^2 \sum_{n=1}^{\infty} \frac{1}{n(n^2+(\ZA)^2)}
\EE
and $r_e=\A/m_e$ is the classical electron radius. As far as total
cross sections are concerned the higher-order contributions tend to a
constant.

Using those results a modification of $\EPEM$ pair creation in
$Z_1+Z_2$ collisions with respect to the lowest order result was
obtained \cite{BertulaniB88}. Such a treatment was not symmetrical
with respect to $Z_1$ and $Z_2$ and an ad hoc symmetrization was
introduced there (see Eq.~(7.3.7) of \cite{BertulaniB88}).

A systematic way to take leading terms of higher order effects into
account in $\EPEM$ pair production is pursued in \cite{IvanovM97}
using Sudakov variables and the impact-factor representation.

Nonperturbative effects are studied also in a light-front approach
\cite{segevW97}. In this approach a gauge transformation on the
time-dependent Dirac equation is performed, in order to remove the
explicit dependence on the long-range part of the interaction. Similar
approaches are also studied in \cite{EichmannRSW98,BaltzM98}.
Numerical evaluation of the non-perturbative effects will be
considered in a future work.

In this context the paper \cite{JackiwKO92} is of interest. In this
work the electromagnetic field of a particle with velocity $v$ is
calculated, see e.g., the textbook result of \cite{JacksonED}. Then
the limit $v\rightarrow c$ is performed. This corresponds to the
electromagnetic fields of a massless particle, which can be regarded
as an ``electromagnetic shock wave''. The results of this paper are
very much reminiscent of the sudden approximation in the semiclassical
theory; for a connection of semiclassical and eikonal methods see also
\cite{Baur91}.

\subsection{Bound-Free Pair Production}

The bound-free pair production, also known as electron-pair production
with capture, is a process, which is also of practical importance in
the collider. It is the process, where a pair is produced but with the
electron not as a free particle, but into an atomic bound state of one
of the nuclei. As this changes the charge state of the nucleus, it is
lost from the beam. Together with the electromagnetic dissociation of
the nuclei (see Sec.~\ref{sec_ga}) these two processes are the
dominant loss processes for heavy ion colliders.

In \cite{BertulaniB88} an approximate value for this cross section is
given as
\BE
\sigma_{capt}^K \approx \frac{33\pi}{10} Z_1^2 Z_2^6 \A^6 r_e^2
\frac{1}{\exp(2\pi Z_2 \A) -1} \left[ \ln\left(\G \delta / 2\right) -
\frac{5}{3}\right] ,
\label{eq_capture}
\EE
where only capture to the $K$-shell is included.
The cross section for all higher shells is expected to be
of the order of 20\% of this cross section (see Eqs 7.6.23 and 24 of
\cite{BertulaniB88}).

The cross section in Eq.~(\ref{eq_capture}) is of the form 
\BE
\sigma= A \ln \G + B.
\label{eq_lnAB}
\EE
This form has been found to be a universal one at
sufficient high values of $\G$. The constant $A$ and $B$ then only
depend on the type of the target.

The above cross section was found making use of the EPA and also using
approximate wave function for bound state and continuum. More precise
calculations exist 
\cite{BaltzRW92,BaltzRW93,BeckerGS87,AsteHT94,AggerS97,RhoadesBrownBS89}
in the literature. Recent calculations within PWBA for high values of
$\G$ have shown that the exact first order results do not differ
significantly from EPA results \cite{MeierHHT98,BertulaniB98}.
Parameterizations for $A$ and $B$\cite{BaltzRW93,AsteHT94} for typical
cases are given in Table~\ref{tab_capture}.
%
% Parameterization of A \ln gamma + B for capture
%
\begin{table}[tbhp]
\begin{center}
\begin{tabular}{|c|c|c|c|c|}
\hline
Ion & $A$ & $B$ & $\sigma(Au,\G=106)$ & $\sigma(Pb,\G=2950)$ \\
\hline
Pb & $15.4$b & $-39.0$b       & 115 b     & 222 b \\
Au & $12.1$b & $-30.7$b       &  90 b     & 173 b \\
Ca & $1.95$mb & $-5.19$mb     & 14 mb     & 27.8 mb\\
O  & $4.50\mu$b & $-12.0\mu$b & 32 $\mu$b & 64.3 $\mu$b \\
\hline
\end{tabular}
\end{center}
\caption{\it
Parameters $A$ and $B$ (see Eq.~(\ref{eq_lnAB})) 
as well as total cross sections for the bound-free pair production for
RHIC and LHC. The parameters are taken from 
\protect\cite{AsteHT94}.}
\label{tab_capture}
\end{table}

For a long time the effect of higher order and nonperturbative
processes has been under investigation. At lower energies, in the
region of few GeV per nucleon, coupled channel calculations have
indicated for a long time, that these give large contributions,
especially at small impact parameters. Newer calculation tend to
predict considerably smaller values, of the order of the first order
result and in a recent article Baltz \cite{Baltz97} finds in the limit
$\G\rightarrow \infty$ that contributions from higher orders are even
slightly smaller than the first order results.

The bound-free pair production was measured in two recent experiments
at the SPS, at $\G=168$ \cite{Krause98} and at $\G \approx 2$
\cite{Belkacem93,Belkacem94}. Both experiments found good agreement
between measurement and calculations.

A similar process was recently used at LEAR (CERN) to produce
antihydrogen. An antiproton beam with a momentum of 1.94~GeV$/c$ hit a
Xenon target ($Z=54$) to produce and detect antihydrogen in the bound
free pair production mechanism\cite{BaurO96}. The same technique is also
used at Fermilab \cite{Blanford98}, where it is also planned to
measure the Lamb shift in antihydrogen as a test of CPT invariance
\cite{MungerBS93,MungerBS94}. Their results are in good agreement with
the recent calculations \cite{MeierHHT98,BertulaniB98}.

We note that electron and positron can also form a bound state,
positronium. This is in analogy to the $\GG$-production of mesons
($q\bar q$ states) discussed in Sec.~\ref{chap_proc}. 
With the known width of the parapositronium $\Gamma((\EPEM)_{n=1}
{}^1S_0 \rightarrow \GG) = m c^2 \A^5 /2$, the photon-photon
production of this bound state was calculated in \cite{Baur90b}.  The
production of orthopositronium, $n=1 {}^3S_1$ was calculated recently
\cite{Ginzburg97}. 

As discussed in Sec.~\ref{chap_proc} the production of
orthopositronium is only suppressed by the factor $(\ZA)^2$, which is
not very small. Therefore one expects that both kind of positronium
are produced in similar numbers. Detailed calculation show that the
three-photon process is indeed not much smaller than the two-photon
process \cite{Ginzburg97,Gevorkyan98}.

%%%%%%%%%%%%%%%%%%%%%%%%%%%%%%%%%%%%%%%%%%%%%%%%%%%%%%%%%%%%%%%%%%%%%%

\section{Conclusion}

In this review the basic properties of electromagnetic processes in
very peripheral hadron-hadron collisions (we deal mainly with
nucleus-nucleus collisions, but proton proton collisions are also
treated) are described. The method of equivalent photons is a well
established tool to describe these kind of reactions. Reliable results
of quasireal photon fluxes and $\GG$-luminosities are
available. Unlike electrons and positrons heavy ions are particles
with an internal structure. We have described how to treat effects
arising from this structure,and we conclude that such effects are well
under control. A problem, which is difficult to judge quantitatively
at the moment, is the influence of strong interactions in grazing
collisions, i.e., effects arising from the nuclear stratosphere and
Pomeron interactions.

The high photon fluxes open up possibilities for photon-photon as well
as photon-nucleus interaction studies up to energies hitherto
unexplored at the forthcoming colliders RHIC and LHC.  Interesting
physics can be explored at the high invariant $\GG$-masses,where
detecting new particles could be within range. Also very interesting
studies within the standard model, i.e., mainly QCD studies will be
possible. This ranges from the study of the total $\GG$-cross section
into hadronic final states up to invariant masses of about 100~GeV to
the spectroscopy of light and heavy mesons.

We also reviewed dilepton production in very peripheral
collisions. This is essentially well understood and gives rise to an
experimental background. Multiple pair production is a strong field
effect of principle interest. Pair production with capture is, in
addition to nuclear fragmentation (sometimes called
``Weizs\"acker-Williams process''), a source of beam loss in the
collider operation. 
   
The future is coming soon. RHIC will be operational in only one year,
LHC in approximately seven years. Therefore the planning of the
experiments and necessary detectors for this kind of physics has to be
done now.  With the new data and new insights, that will come from
these experiments, new work and theoretical understanding will be
required. As an ancient motto says: ``No surprise would be a
surprise''. 

\section{Acknowledgments}

We have benefited from numerous discussions with many people over the
past years. We thank all of them, especially we wish to mention
A.~Alscher, N.~Baron, C.~A.~Bertulani, I.~Ginzburg, O.~Conradt,
S.~Datz, K.~Eggert, R.~Engel, S.~Gevorkyan, M.~Greiner, Z.~Halabuka,
S.~Klein, F.~Krauss, H.~Meier, N.~N.~Nikolaev, S.~Sadovsky, R.~Schuch,
and G.~Soff.

%%%%%%%%%%%%%%%%%%%%%%%%%%%%%%%%%%%%%%%%%%%%%%%%%%%%%%%%%%%%%%%%%%%%%%


\begin{thebibliography}{100}

\bibitem{BertulaniB88}
C.~A. Bertulani and G. Baur, Phys. Rep. {\bf 163},  299  (1988).

\bibitem{BaurR94}
G. Baur and H. Rebel, J. Phys.~G {\bf 20},  1  (1994).

\bibitem{BaurR96}
G. Baur and H. Rebel, Annu. Rev. Nucl. Part. Sci. {\bf 46},  321  (1996).

\bibitem{Primakoff51}
H. Primakoff, Phys. Rev. {\bf 31},  899  (1951).

\bibitem{Moshammer97}
R. Moshammer {\it et~al.}, Phys. Rev. Lett. {\bf 79},  3621  (1997).

\bibitem{LandauL34}
L.~D. Landau and E.~M. Lifshitz, Phys. Z. Sowjet. {\bf 6},  244  (1934).

\bibitem{Soff80}
G. Soff,  in {\em Proc. 18th Winter School (Bielsko-Biala, Poland), 1980},
  edited by A. Balanda and Z. Stachura (~, ~, 1980), p.\ 201.

\bibitem{AlexanderGM87}
G. Alexander, E. Gotsman, and U. Maor, Z. Phys. C {\bf 34},  329  (1987).

\bibitem{BaurB88}
G. Baur and C.~A. Bertulani, Z. Phys. A {\bf 330},  77  (1988).

\bibitem{GrabiakMG89}
M. Grabiak, B. M{\"u}ller, W. Greiner, and P. Koch, J. Phys.~G {\bf 15},  L25
  (1989).

\bibitem{Papageorgiu89}
E. Papageorgiu, Phys. Rev.~D {\bf 40},  92  (1989).

\bibitem{Baur90d}
G. Baur,  in {\em CBPF Int. Workshop on relativistic aspects of nuclear
  physics, Rio de Janeiro, Brazil 1989}, edited by T. Kodama {\it et~al.}
  (World Scientific, Singapore, 1990), p.\ 127.

\bibitem{BaurF90}
G. Baur and L.~G. {Ferreira Filho}, Nucl. Phys.~A {\bf 518},  786  (1990).

\bibitem{CahnJ90}
N. Cahn and J.~D. Jackson, Phys. Rev.~D {\bf 42},  3690  (1990).

\bibitem{VidovicGB93}
M. Vidovi{\'c}, M. Greiner, C. Best, and G. Soff, Phys. Rev.~C {\bf 47},  2308
  (1993).

\bibitem{KraussGS97}
F. Krauss, M. Greiner, and G. Soff, Prog. Part. Nucl. Phys. {\bf 39},  503
  (1997).

\bibitem{Vannucci80}
F. Vannucci,  in {\em $\gamma\gamma$ Collisions, Proceedings, Amiens 1980},
  Vol.~134 of {\em Lecture Notes in Physics}, edited by G. Cochard (Springer,
  Heidelberg, Berlin, New York, 1980).

\bibitem{KleinS97a}
S. Klein and E. Scannapieco,  in {\em Photon '97, Egmond aan Zee}, edited by
  (World Scientific, Singapore, 1997), p.\ 369.

\bibitem{KleinS97b}
S. Klein and E. Scannapieco, Coherent Photons and Pomerons in Heavy Ion
  Collisions, presented at 6th Conference on the Intersections of Particle and
  Nuclear Physics, May 1997, Big Sky, Montana, STAR Note 298, LBNL-40495, 1997.

\bibitem{KleinS95a}
S. Klein and E. Scannapieco, STAR Note 243, 1995.

\bibitem{KleinS95b}
S. Klein,  in {\em Photon '95, Sheffield}, edited by D.~J. Miller, S.~L.
  Cartwright, and V. Khoze (World Scientific, Singapore, 1995), p.\ 417.

\bibitem{HenckenKKS96}
{K. Hencken, {Yu.} V. Kharlov, G. V. Khaustov, S. A. Sadovsky, and V. D.
  Samoylenko}, TPHIC, event generator of two photon interactions in heavy ion
  collisions, IHEP-96-38, 1996.

\bibitem{Felix97}
K. Eggert {\it et~al.}, FELIX Letter of Intent, CERN/LHCC 97--45, LHCC/I10,
  1997.

\bibitem{BaurHTS98}
{G. Baur, K. Hencken, D. Trautmann, S. Sadovsky, and Yu. Kharlov},
  Photon-Photon Physics with heavy ions at CMS, CMS Note 1998/009, available
  from the CMS information server at http://cmsserver.cern.ch, 1998.

\bibitem{BlockC85}
M.~M. Block and R. Cahn, Rev. Mod. Phys. {\bf 57},  563  (1985).

\bibitem{Abe94a}
F. Abe {\it et~al.}, Phys. Rev.~D {\bf 50},  5550  (1994).

\bibitem{Abe94b}
F. Abe {\it et~al.}, Phys. Rev.~D {\bf 50},  5535  (1994).

\bibitem{Abe94c}
F. Abe {\it et~al.}, Phys. Rev.~D {\bf 50},  5518  (1994).

\bibitem{Greider65}
K.~R. Greider, Annu. Rev. Nucl. Part. Sci. {\bf 15},  291  (1965).

\bibitem{JacksonED}
J.~D. Jackson, {\em Classical Electrodynamics} (John Wiley, New York, 1975).

\bibitem{NoerenbergW80}
W. N{\"o}renberg and H.~A. Weidenm{\"u}ller, {\em Introduction to the theory of
  heavy ion collisions} (Springer, Berlin, Heidelberg, New York, 1980).

\bibitem{Frahn72}
W.~E. Frahn, Ann. Phys. {\bf 72},  524  (1972).

\bibitem{PDG96}
R.~M. Barnett {\it et~al.}, Phys. Rev.~D {\bf 54},  1  (1996).

\bibitem{KopeliovichNP88}
B.~Z. Kopeliovich, N.~N. Nikolaev, and I.~K. Potashnikova, Phys. Rev.~D {\bf
  39},  769  (1988).

\bibitem{Glauber67}
R.~J. Glauber, {\em High Energy Physics and Nuclear Structure} (North-Holland,
  Amsterdam, 1967), p.\ 311.

\bibitem{EngelGLS92}
J. Engel, T.~K. Gaisser, P. Lipari, and T. Stanev, Phys. Rev.~D {\bf 46},  5013
   (1992).

\bibitem{BertschBS89}
G. Bertsch, B.~A. Brown, and H. Sagawa, Phys. Rev.~C {\bf 39},  1154  (1989).

\bibitem{Abbiendi97}
G. Abbiendi, Nuovo Cim. {\bf 20},  1  (1997).

\bibitem{Engel97}
R. Engel, Ph.D. thesis, Universit{\"a}t Gesamthochschule Siegen, 1997.

\bibitem{BudnevGM75}
V.~M. Budnev, I.~F. Ginzburg, G.~V. Meledin, and V.~G. Serbo, Phys. Rep. {\bf
  15},  181  (1975).

\bibitem{WintherA79}
A. Winther and K. Alder, Nucl. Phys.~A {\bf 319},  518  (1979).

\bibitem{AbramowitzS64}
M. Abramowitz and I.~A. Stegun, {\em Handbook of mathematical Functions}
  (Dover, New York, 1965).

\bibitem{BaurF91}
G. Baur and L.~G. {Ferreira Filho}, Phys. Lett.~B {\bf 254},  30  (1991).

\bibitem{BaurB93}
G. Baur and N. Baron, Nucl. Phys.~A {\bf 561},  628  (1993).

\bibitem{Baur92}
G. Baur, Z. Phys. C {\bf 54},  419  (1992).

\bibitem{BaronB94}
N. Baron and G. Baur, Phys. Rev.~C {\bf 49},  1127  (1994).

\bibitem{VidovicGS93}
M. Vidovi{\'c}, M. Greiner, and G. Soff, Phys. Rev.~C {\bf 48},  2011  (1993).

\bibitem{BaltzRW96}
A.~J. Baltz, M.~J. Rhoades-Brown, and J. Weneser, Phys. Rev. E {\bf 54},  4233
  (1996).

\bibitem{BaltzS98}
A.~J. Baltz and M. Strikman, Phys. Rev.~D {\bf 57},  548  (1998).

\bibitem{Baltz98}
A. Baltz, C. Chasman, and S.~N. White, nucl-ex/9801002, 1998.

\bibitem{HenckenTB96}
K. Hencken, D. Trautmann, and G. Baur, Phys. Rev.~C {\bf 53},  2532  (1996).

\bibitem{HenckenTB95}
K. Hencken, D. Trautmann, and G. Baur, Z. Phys. C {\bf 68},  473  (1995).

\bibitem{Conradt96}
O. Conradt, Diplomarbeit, Universit{\"a}t Basel, (unpublished), 1996.

\bibitem{Kniehl91}
B. Kniehl, Phys. Lett.~B {\bf 254},  267  (1991).

\bibitem{Chanfray93}
G. Chanfray, J. Delorme, M. Ericson, and A. Molinari, Nucl. Phys.~A {\bf 556},
  439  (1993).

\bibitem{Baron94}
N. Baron, Ph.D. thesis, Forschungszentrum J{\"u}lich, Institut f{\"u}r
  Kernphysik J{\"u}l-2846, 1994.

\bibitem{deForestW66}
T. deForest and J.~D. Walecka, Adv. Phys. {\bf 15},  1  (1966).

\bibitem{HalzenM84}
F. Halzen and A.~D. Martin, {\em Quarks \& Leptons} (John Wiley \& Sons, New
  York, 1984).

\bibitem{DreesGN94}
M. Drees, R.~M. Godbole, N. Nowakowski, and S.~D. Rindami, Phys. Rev.~D {\bf
  50},  2335  (1994).

\bibitem{OhnemusWZ94}
J. Ohnemus, T.~F. Walsh, and P.~M. Zerwas, Phys. Lett.~B {\bf 328},  369
  (1994).

\bibitem{Martin93a}
A.~D. Martin {\it et~al.}, Phys. Lett.~B {\bf 306},  306  (1993).

\bibitem{Martin93b}
A.~D. Martin {\it et~al.}, Phys. Lett.~B {\bf 309},  492  (1993).

\bibitem{BauerSYP78}
T.~H. Bauer, R.~D. Spital, D.~R. Yennie, and F.~M. Pipkin, Rev. Mod. Phys. {\bf
  50},  261  (1978).

\bibitem{NorburyW98}
J.~W. Norbury and M. Waldsmith, Phys. Rev.~C {\bf 57},  1525  (1998).

\bibitem{BrechtmannH88a}
C. Brechtmann and W. Heinrich, Z. Phys. A {\bf 330},  407  (1988).

\bibitem{BrechtmannH88b}
C. Brechtmann and W. Heinrich, Z. Phys. A {\bf 331},  463  (1988).

\bibitem{PriceGW88}
P.~B. Price, R. Guaxiao, and W.~T. Williams, Phys. Rev. Lett. {\bf 61},  2193
  (1988).

\bibitem{Pshenichnov98}
I.~A. Pshenichnov {\it et~al.}, Phys. Rev.~C {\bf 57},  1920  (1998).

\bibitem{BaurB89}
G. Baur and C.~A. Bertulani, Nucl. Phys.~A {\bf 505},  835  (1989).

\bibitem{Datz97}
S. Datz {\it et~al.}, Phys. Rev. Lett. {\bf 79},  3355  (1997).

\bibitem{NorburyB93}
J.~W. Norbury and G. Baur, Phys. Rev.~C {\bf 48},  1915  (1993).

\bibitem{Crittenden97}
J.~A. Crittenden, {\em Exclusive production of neutral vector mesons at the
  electron-proton collider HERA}, Vol.~140 of {\em Springer tracts in modern
  physics} (Springer, Heidelberg, 1997).

\bibitem{RyskinRML97}
M.~G. Ryskin, R.~G. Roberts, A.~D. Martin, and E.~M. Levin, Z. Phys. C {\bf
  76},  231  (1997).

\bibitem{Breitweg97}
J. Breitweg {\it et~al.}, Z. Phys. C {\bf 76},  599  (1997).

\bibitem{Miller98}
D.~J. Miller, J. Phys.~G {\bf 24},  317  (1998).

\bibitem{FreedmanST77}
D.~Z. Freedman, D.~N. Schramm, and D.~L. Tubbs, Annu. Rev. Nucl. Part. Sci.
  {\bf 27},  167  (1977).

\bibitem{Ginzburg97}
I.~F. Ginzburg, private communication.

\bibitem{Ginzburg87}
I.~F. Ginzburg {\it et~al.}, Nucl. Phys.~B {\bf 284},  685  (1987).

\bibitem{Ginzburg88}
I.~F. Ginzburg {\it et~al.}, Nucl. Phys.~B {\bf 296},  569  (1988).

\bibitem{Ginzburg92}
I.~F. Ginzburg {\it et~al.}, Nucl. Phys.~B {\bf 388},  376  (1992).

\bibitem{Gevorkyan98}
S.~R. Gevorkyan {\it et~al.}, hep-ph9804264.

\bibitem{HofmannSSG91}
C. Hofmann, G. Soff, A. Sch{\"a}fer, and W. Greiner, Phys. Lett.~B {\bf 262},
  210  (1991).

\bibitem{BaronB93}
N. Baron and G. Baur, Phys. Rev.~C {\bf 48},  1999  (1993).

\bibitem{GreinerVHS95}
M. Greiner {\it et~al.}, Phys. Rev.~C {\bf 51},  911  (1995).

\bibitem{Weinberg97}
S. Weinberg, {\em The Quantum Theory of Fields} (Cambridge University Press,
  Cambridge, 1997), Vol.~1.

\bibitem{KolanoskiZ88}
H. Kolanoski and P. Zerwas,  in {\em High Energy Electron-Positron Physics},
  edited by A. Ali and P. S{\"o}ding (World Scientific, Singapore, 1988).

\bibitem{BergerW87}
{Ch. Berger and W. Wagner}, Phys. Rep. {\bf 176C},  1  (2987).

\bibitem{Amiens80}
{\em {$\gamma\gamma$} Collisions, Proceedings, Amiens 1980}, Vol.~134 of {\em
  Lecture Notes in Physics}, edited by G. Cochard and P. Kessler (Springer,
  Berlin, 1980).

\bibitem{SanDiego92}
{\em Proc. 9th International Workshop on Photon-Photon Collisions, San Diego
  (1992)} (World Scientific, Singapore, 1992).

\bibitem{Sheffield95}
{\em Photon'95, Xth International Workshop on Gamma-Gamma Collisions and
  related Processes}, edited by D.~J. Miller, S.~L. Cartwright, and V. Khoze
  (World Scientific, Singapore, 1995).

\bibitem{Egmond97}
{\em Photon'97, XIth International Workshop on Gamma-Gamma Collisions and
  related Processes, Egmond aan Zee}, edited by A. Buijs (World Scientific,
  Singapore, 1997).

\bibitem{L3:97}
{L3 collaboration}, Phys. Lett.~B {\bf 408},  450  (1997).

\bibitem{Yang48}
C.~N. Yang, Phys. Rev. {\bf 77},  242  (1948).

\bibitem{Cooper88}
S. Cooper, Annu. Rev. Nucl. Part. Sci. {\bf 28},  705  (1988).

\bibitem{Cartwright98}
S. Cartwright {\it et~al.}, J. Phys.~G {\bf 24},  457  (1998).

\bibitem{Godang97}
R. Godang {\it et~al.}, Phys. Rev. Lett. {\bf 79},  3829  (1997).

\bibitem{Telnov95}
V. Telnov,  in {\em Photon '95, Sheffield}, edited by D.~J. Miller, S.~L.
  Cartwright, and V. Khoze (World Scientific, Singapore, 1995), p.\ 369.

\bibitem{ginzburg95}
I.~F. Ginzburg,  in {\em Photon '95, Sheffield}, edited by D.~J. Miller, S.~L.
  Cartwright, and V. Khoze (World Scientific, Singapore, 1995), p.\ 399.

\bibitem{DreesEZ89}
M. Drees, H. Ellis, and D. Zeppenfeld, Phys. Lett.~B {\bf 223},  454  (1989).

\bibitem{MuellerS90}
B. {M\"uller} and A.~J. Schramm, Phys. Rev.~D {\bf 42},  3699  (1990).

\bibitem{Papageorgiu95}
E. Papageorgiu, Phys. Lett.~B {\bf 352},  394  (1995).

\bibitem{Norbury90}
J. Norbury, Phys. Rev.~D {\bf 42},  3696  (1990).

\bibitem{ChoudhuryK97}
D. Choudhury and M. Krawczyk, Phys. Rev.~D {\bf 55},  2774  (1997).

\bibitem{Renard83}
F.~M. Renard, Phys. Lett.~B {\bf 126},  59  (1983).

\bibitem{BaurFF84}
U. Baur, H. Fritzsch, and H. Faissner, Phys. Lett.~B {\bf 135},  313  (1984).

\bibitem{herafuture96}
 in {\em Future Physics at HERA}, edited by G. Ingelman, A. {De Roeck}, and R.
  Klanner (DESY, Hamburg, 1996).

\bibitem{Mueller98}
A.~H. M{\"u}ller, Eur. Phys. J. A {\bf 1},  19  (1998).

\bibitem{EngelRR97}
R. Engel {\it et~al.}, Z. Phys. C {\bf 74},  687  (1997).

\bibitem{Engel98}
R. Engel, private communication.

\bibitem{MuellerS91}
B. {M\"uller} and A.~J. Schramm, Nucl. Phys.~A {\bf 523},  677  (1991).

\bibitem{SchrammR97}
A.~J. Schramm and D.~H. Reeves, Phys. Rev.~D {\bf 55},  7312  (1997).

\bibitem{NystrandK97}
J. Nystrand and S. Klein, Two Photons Physics at RHIC: Separating Signals from
  Backgrounds, talk presented at ``Hadron'97'', Brookhaven National Laboratory,
  August 1997, STAR Note 315, LBNL-41111 Nov.97, 1997.

\bibitem{Griffin98}
J.~J. Griffin, The APEX/EPOS Quandary: the way out via low Energy Studies,
  nucl-th/9802044, 1998.

\bibitem{EichlerM95}
J. Eichler and W.~E. Meyerhof, {\em Relativistic Atomic Collisions} (Academic
  Press, San Diego, 1995).

\bibitem{Baur90}
G. Baur, Phys. Rev.~A {\bf 42},  5736  (1990).

\bibitem{heitler34}
W. Heitler, {\em The Quantum Theory of Radiation} (Oxford University Press,
  London, 1954).

\bibitem{Baur90c}
G. Baur, Phys. Rev.~D {\bf 41},  3535  (1990).

\bibitem{BestGS92}
C. Best, W. Greiner, and G. Soff, Phys. Rev.~A {\bf 46},  261  (1992).

\bibitem{RhoadesBrownW91}
M.~J. Rhoades-Brown and J. Weneser, Phys. Rev.~A {\bf 44},  330  (1991).

\bibitem{HenckenTB95a}
K. Hencken, D. Trautmann, and G. Baur, Phys. Rev.~A {\bf 51},  998  (1995).

\bibitem{HenckenTB95b}
K. Hencken, D. Trautmann, and G. Baur, Phys. Rev.~A {\bf 51},  1874  (1995).

\bibitem{Guclu95}
M.~C. G{\"u\c cl\"u} {\it et~al.}, Phys. Rev.~A {\bf 51},  1836  (1995).

\bibitem{Bottcher89}
C. Bottcher and M.~R. Strayer, Phys. Rev.~D {\bf 39},  1330  (1989).

\bibitem{alscherHT97}
A. Alscher, K. Hencken, D. Trautmann, and G. Baur, Phys. Rev.~A {\bf 55},  396
  (1997).

\bibitem{BaronB92}
N. Baron and G. Baur, Phys. Rev.~D {\bf 46},  R3695  (1992).

\bibitem{BaurB93b}
G. Baur and N. Baron, Z. Phys. C {\bf 60},  95  (1993).

\bibitem{Baur92b}
G. Baur, Nucl. Phys.~A {\bf 538},  187c  (1992).

\bibitem{BaronB92b}
N. Baron and G. Baur, Phys. Rev.~D {\bf 46},  4897  (1992).

\bibitem{stekas81}
J. Stekas {\it et~al.}, Phys. Rev. Lett. {\bf 47},  1686  (1981).

\bibitem{BottcherSAE90}
C. Bottcher, M.~R. Strayer, C.~J. Albert, and D.~J. Ernst, Phys. Lett.~B {\bf
  237},  175  (1990).

\bibitem{LandauLQED}
L.~D. Landau and E.~M. Lifschitz, {\em Quantenelektrodynamik}, No.~IV in {\em
  Lehrbuch der theoretischen Physik} (Akademie Verlag, Berlin, 1986).

\bibitem{IvanovM97}
D. Ivanov and K. Melnikov, Phys. Rev.~D {\bf 57},  4025  (1998).

\bibitem{segevW97}
B. Segev and J.~C. Wells, Phys. Rev.~A {\bf 57},  1849  (1998).

\bibitem{EichmannRSW98}
U. Eichmann, J. Reinhardt, S. Schramm, and W. Greiner, nucl-th/9804064.

\bibitem{BaltzM98}
A.~J. Baltz and L. McLerran, nucl-th/9804042.

\bibitem{JackiwKO92}
R. Jackiw, D. Kabat, and M. Ortiz, Phys. Lett.~B {\bf 277},  148  (1992).

\bibitem{Baur91}
G. Baur, Nucl. Phys.~A {\bf 531},  685  (1991).

\bibitem{BaltzRW92}
A.~J. Baltz, M.~J. Rhoades-Brown, and J. Weneser, Phys. Rev.~A {\bf 44},  5569
  (1992).

\bibitem{BaltzRW93}
A.~J. Baltz, M.~J. Rhoades-Brown, and J. Weneser, Phys. Rev.~A {\bf 48},  2002
  (1993).

\bibitem{BeckerGS87}
U. Becker, N. Gr{\"u}n, and W. Scheid, J. Phys.~B {\bf 20},  6563  (1987).

\bibitem{AsteHT94}
A. Aste, K. Hencken, D. Trautmann, and G. Baur, Phys. Rev.~A {\bf 50},  3980
  (1994).

\bibitem{AggerS97}
C.~K. Agger and A.~H. S{\o}rensen, Phys. Rev.~A {\bf 55},  402  (1997).

\bibitem{RhoadesBrownBS89}
M.~J. Rhoades-Brown, C. Bottcher, and M.~R. Strayer, Phys. Rev.~A {\bf 40},
  2831  (1089).

\bibitem{MeierHHT98}
H. Meier {\it et~al.}, Eur. Phys. J. C {\bf in print},  ~  (1998).

\bibitem{BertulaniB98}
C.~A. Bertulani and G. Baur, Phys. Rev.~D {\bf in print},  ~  (1998).

\bibitem{Baltz97}
A. Baltz, Phys. Rev. Lett. {\bf 78},  1231  (1997).

\bibitem{Krause98}
H.~F. Krause {\it et~al.}, Phys. Rev. Lett. {\bf 80},  1190  (1998).

\bibitem{Belkacem93}
A. Belkacem {\it et~al.}, Phys. Rev. Lett. {\bf 71},  1514  (1993).

\bibitem{Belkacem94}
A. Belkacem {\it et~al.}, Phys. Rev. Lett. {\bf 73},  2432  (1994).

\bibitem{BaurO96}
G. Baur {\it et~al.}, Phys. Lett.~B {\bf 368},  251  (1996).

\bibitem{Blanford98}
G. Blanford {\it et~al.}, Phys. Rev. Lett. {\bf 80},  3037  (1998).

\bibitem{MungerBS93}
C.~T. Munger, S.~J. Brodsky, and I. Schmidt, Hyperfine Interactions {\bf 76},
  175  (1993).

\bibitem{MungerBS94}
C.~T. Munger, S.~J. Brodsky, and I. Schmidt, Phys. Rev.~D {\bf 49},  3228
  (1994).

\bibitem{Baur90b}
G. Baur,  in {\em Perspectives on Photon Interactions with Hadrons and Nuclei},
  {\em Lecture Notes in Physics}, edited by M. Schumacher and G. Tamas
  (Springer Verlag, Berlin, Heidelberg, New York, 1990), p.\ 111.

\end{thebibliography}
\end{document}